\documentclass[lettersize,journal]{IEEEtran}
\usepackage{amsmath,amsfonts}
\usepackage{array}
\usepackage{textcomp}
\usepackage{stfloats}
\usepackage{url}
\usepackage{verbatim}
\usepackage{cite}
\usepackage{graphicx}
\usepackage{setspace}
\usepackage{epstopdf}
\usepackage{adjustbox} % for adjusting 
\DeclareGraphicsExtensions{.pdf,.jpeg,.png}
\usepackage{enumitem}
\usepackage[linesnumbered,ruled]{algorithm2e}
\usepackage{amsmath}
\usepackage{lipsum}
\usepackage{mathtools}
\usepackage{cuted}
\usepackage{array}
\usepackage{subfigure}
\usepackage{amssymb}
\usepackage{multirow}
\usepackage{makecell}
\usepackage{algorithmic}
\usepackage{xcolor}
\def\BibTeX{{\rm B\kern-.05em{\sc i\kern-.025em b}\kern-.08em
ThedeltaT\kern-.1667em\lower.7ex\hbox{E}\kern-.125emX}}

\SetCommentSty{mycommfont}
\usepackage{array,multirow,makecell}
\SetKwInput{KwInput}{Input}    
\SetKwInput{KwRepeat}{Repeat} % Set the Input
\SetKwInput{KwOutput}{Output}    
\begin{document}

\title{Meta-Learning Driven Movable-Antenna-assisted Full-Duplex RSMA for Multi-User Communication}

\author{Shreya Khisa, Ali Amhaz, Mohamed Elhattab, Chadi Assi, Sanaa Sharafeddine
\thanks{}% <-this % stops a space
\thanks{}}

%S\markboth{Journal of \LaTeX\ Class Files,~Vol.~14, No.~8, August~2021}%%{Shell \MakeLowercase{\textit{et al.}}: A Sample Article Using IEEEtran.cls for IEEE Journals}


\maketitle

\begin{abstract}
Full-duplex (FD) radios at base stations (BS)s have gained significant interest because of their ability to simultaneously transmit and receive signals on the same frequency band. However, FD communication is hindered by self-interference (SI) and intra-cell interference caused by simultaneous uplink (UL) transmissions affecting downlink (DL) reception. Movable antenna (MA) technology has recently emerged as a promising solution to mitigate interference by dynamically adjusting antenna positions within the transmitter or receiver region, enhancing multiplexing gain and spectral efficiency (SE). In traditional FD systems with fixed-position antennas (FPAs), residual SI and intra-cell interference are mitigated either through beamforming and power allocation or by leveraging the inherent interference management strategy of multiple access schemes like rate-splitting multiple access (RSMA). Utilizing MA
in FD systems adds an extra degree of freedom to effectively
mitigate interference in conjunction with other strategies. In this context, we integrate MA with FD systems to evaluate and investigate its effectiveness in handling SI and intra-cell interference. Moreover, we utilize RSMA as our multiple access technique in both UL and DL transmission. To achieve the full potential of the system, we evaluate three different scenarios with FD-BS-RSMA with MAs where our goal is to maximize the total sum rate of the system by jointly optimizing the transmitting and receiving beamforming vectors, UL user equipment (UE) transmission power, MA positions, and common stream split ratio of RSMA while satisfying the minimum data rate requirements of all UEs, common stream rate constraint, power budget requirements of BS and UL UEs, and inter-MA distance. The formulated optimization problem is highly non-convex in nature, and hence, we propose a gradient-based meta-learning (GML) approach, which can handle the non-convexity in a discrete manner by optimizing each variable in a different neural network. Our simulation results show that introducing MA with FD-BS-RSMA  system not only improves the overall SE but also effectively tackles SI and intra-cell interference. Our proposed three scenarios, which integrate MA with the FD-BS-RSMA system, achieve performance improvements ranging from 41\% to 240\% compared to traditional FPA systems. 
\end{abstract}

\begin{IEEEkeywords}
Movable antenna, RSMA, full-duplex (FD), uplink, downlink, meta-learning
\end{IEEEkeywords}

\section{Introduction}
%\IEEEPARstart{A}{ccording} 
With the emergence of sixth-generation (6G) mobile communication systems, there is a pressing demand to explore communication technologies that offer high spectral efficiency (SE), reliability, energy efficiency and latency \cite{10258345}. In addition, the continuous surge in data-intensive applications places highly demanding requirements on wireless communication systems. However, due to limited spectrum resources, enhancing the communication rate using traditional half-duplex (HD) systems remains a challenge. In order to fulfill these stringent requirements and challenges, several promising technologies have emerged, such as millimeter wave (mmWave) communication, massive multiple-input multiple output (mMIMO) and full-duplex (FD) communications \cite{10258345}. Among these technologies, FD communication has garnered significant interest because of its superior SE and co-time co-frequency transceiving capability compared to its HD counterpart \cite{10315060, 10158724}.

In FD communications, multiple terminals can simultaneously transmit and receive information within the same frequency band \cite{ding2024movable}. In particular, FD communication is capable of doubling up the SE by accommodating both uplink (UL) and downlink (DL) communications at the same time-frequency resource. However, allocating transmit and receive signals to operate in the same frequency and time slot inherently leads to significant interference issues. Specifically, the most critical issue is residual self-interference (SI) from the transmit antennas of the base station (BS) to the receive antennas, which could be much higher than the received signal. In addition, residual SI and DL user equipment (UE) experience significant intra-cell interference caused by UL UE transmissions, which greatly limits the potential gains
of using FD technology. Hence, FD technology will greatly suffer quite a loss if the interferences are not handled in an efficient manner.

Recently, movable antenna (MA) has emerged as a contender technology that is capable of effectively reducing interference in the MIMO system by positioning each MA in the transmitter/receiver region to shift from areas of high interference to locations with minimal interference, resulting in higher multiplexing gain and higher SE \cite{zhu2023movable}. Moreover, by moving the antenna positions in a certain region, MA adds a new degree of freedom to exploit the benefit of the spatial domain. Specifically, unlike traditional fixed-position antennas (FPAs), MAs can be dynamically adjusted within a spatial region to optimize channel conditions and enhance communication performance. In particular, an MA is linked to the radio frequency (RF) chain via a flexible cable, enabling it to be repositioned within a spatial region with the help of a driver.   Although traditional MIMO systems using antenna selection can achieve spatial degrees of freedom, they generally require a large number of antennas to cover the entire spatial region and fully leverage spatial diversity and multiplexing gains \cite{zhu2023movable}. In contrast, a much smaller number of MAs, even a single MA, can fully exploit these degrees of freedom because of their flexible positioning.

Another crucial aspect of enhancing SE lies in the careful selection of multiple access techniques. In the past couple of years, Rate-Splitting Multiple Access (RSMA) has recently emerged as a promising non-orthogonal multiple access technique, offering enhanced flexibility in interference management for next-generation wireless communications \cite{mao2022rate}. The core concept of RSMA is to handle multi-user interference by partially decoding it and partially treating it as noise \cite{mao2022rate}.
In RSMA, the BS divides the UE signals into common and private components and transmits the combined signal using superposition coding (SC). At the receiver end, the common stream is first decoded while considering all private streams as interference. Once decoded, the common stream is removed from the received signal through successive interference cancellation (SIC). Subsequently, each UE decodes its private stream while treating private streams from other UEs as interference.
This adaptive interference management strategy enables RSMA to serve as an intermediate approach between space-division multiple access (SDMA), which entirely treats multi-user interference as noise, and non-orthogonal multiple access (NOMA), which fully decodes interference.

Considering the benefits of these technologies and the challenges in optimizing resource allocation, extensive research has been performed to evaluate their performance by developing system models that incorporate MAs, FD, and RSMA in different scenarios, which also analyze their interdependencies. The following subsection will present a comprehensive review of the latest advancements in these areas.
\subsection{State-of-the-art}
\subsubsection{FD-BS-based wireless systems}
Motivated by the substantial gain brought by the FD communication system, several works have been reported in the literature investigating the interaction between FD and other emerging technologies. For example, the authors in \cite{10315060} proposed a multi-cell FD system incorporating reconfigurable intelligent surface (RIS) where they showed that the system could significantly improve cell edge users performance. Meanwhile, the authors in \cite{10477869} presented a real-world evaluation of the FD system with mmWave communication and measured the impact of the SI channel. The paper in \cite{10770280} explored the optimization of long-term energy efficiency  for the FD cell-free massive
MIMO system and the formulated problem was solved utilizing a reinforcement learning-based approach.  Moving on, the authors in \cite{9448430} studied the effects of joint transmit
and receive antenna selection problems on FD MIMO networks’ performance. In this context, they formulated a sum-rate maximization problem, and the problem was solved using a generalized Bender’s decomposition-based
algorithm. On the other hand, the authors in \cite{park2022joint} studied a joint antenna and Internet
of Things device scheduling problem for FD MIMO wireless-powered communication network over time-varying fading channels.
First, they formulated an optimization problem to maximize the average sum rate of Internet of Things devices while satisfying their minimum average data
rate requirements, and then they solved the problem by proposing a scheduling algorithm based on
Lagrangian duality and the stochastic optimization theory. 
\subsubsection{RSMA/NOMA-assisted FD-BS systems}
Inspired by the potential of RSMA/NOMA and FD, several research works have been performed incorporating these two technologies in different scenarios. 
The authors in \cite{allu2024robust} investigated an FD-integrated RSMA
scheme for improved SE and energy efficiency performance and compared it to the conventional
power-domain schemes. In this regard, the authors formulated a multi-objective optimization problem that aimed to jointly maximize
energy efficiency and SE by jointly optimizing
transmit power budget 
and minimum rate under the assumption of a channel state information error
model. Then, the multi-objective  problem was converted
into a single-objective optimization  problem using the
weighted sum method, and successive convex approximation (SCA)
and the S-procedure was utilized to solve the formulated problem. 
Another work presented a model consisting of FD RIS and NOMA with the consideration of hardware impairment \cite{mead2024hardware}.  In this regard, they proposed an iterative algorithm that
optimized the transmit power at the BS and UL users and the
passive beamforming at the RIS.  An RSMA FD-based vehicular-to-everything network has been presented in \cite{10663924}, which is also aided by RIS to maximize the SE. The formulated non-convex problem is solved utilizing a deep reinforcement learning-based optimization algorithm. 
Another work has presented a model consisting of FD and RIS and NOMA with the consideration of hardware impairment \cite{mead2024hardware}.  In this regard, they proposed an iterative algorithm that
optimizes the transmit power at the BS and UL users and the
passive beamforming at the RIS. 
\subsubsection{MA-enabled MISO/MIMO/FD systems}
MA technology was first introduced in \cite{10318061}, where a thorough study was conducted to highlight its promising potential compared to traditional FPA systems, which suffer from limited flexibility in the spatial domain. Since then, researchers have been exploring innovative system models incorporating various technologies to further amplify its advantages. Multiple-input single-output (MISO)/MIMO technologies, cutting-edge advancements in 5G networks, and key enablers for next-generation wireless systems have proven to increase data throughput and enhance system capacity. As a result, several studies have shifted focus toward MA-enabled MISO/MIMO communication systems. The authors in \cite{10243545} examined a MIMO system empowered by MA technology at the transmitter's and receiver's sides. To characterize the capacity in the model, an optimization problem was formulated to maximize the sum rate by jointly determining the covariance of the transmit signal along with the MAs positions. %In \cite{10851455}, a multi-user hybrid beamforming system enabled by MA technology was studied to maximize the achievable sum rate. In this regard, the authors formulated an optimization problem that jointly optimizes the MA positions in addition to the digital and analog beamforming vectors. Then, to address the presented problem and tackle its non-concave/non-convex structure, an alternating optimization (AO)  approach was deployed to produce an efficient solution. 
Besides, in \cite{10508218}, the authors introduced a graph-based technique for optimizing MA positions, contrasting with conventional optimization methods. Their approach involved dividing the transmit area into discrete points, converting the continuous optimization of antenna positions into a problem of selecting the best sampling points based on local channel data. Moving on, the authors in \cite{ding2024movable} studied an FD-based MA system focusing on the perspective of physical layer security.  Particularly, in their model, a BS equipped with MAs operates in FD mode and transmits artificial noise (AN) to simultaneously protect UL and DL transmission. Then, they formulated a sum of secrecy rates maximization problem by jointly optimizing MA positions, the transmit, receive, and
AN beamformers at the BS, and the UL powers. Finally, the optimization problem was solved utilizing the alternating optimization (AO) algorithm by decomposing the main problem into multiple sub-problems. For the optimization of beamforming vectors and UL power, SCA was utilized, and for MA position optimization, a particle swarm-based approach was proposed. Meanwhile, the authors in \cite{guo2024movable}, examined an FD-integrated sensing and communication-based system with MAs to promote
communication capability with guaranteed sensing performance
via jointly designing beamforming, power allocation, and receiving
filters, and MA configuration towards maximizing the sum rate. The optimization problem was later solved using the majorization-minimization method. 
\subsubsection{RSMA-based MA-enabled wireless systems}
In order to enhance connectivity and SE in upcoming wireless networks, various studies have begun to concentrate on strengthening MA-enabled networks using non-orthogonal transmission methods, particularly NOMA and RSMA.
The authors of \cite{zhang2024sumratemaximizationmovable} analyzed an RSMA-equipped model in which the BS is equipped with a set of linear MAs. Motivated by the advantages of MA technology in terms of efficient exploitation of spatial diversity, they formulated an optimization problem to maximize the achievable sum rate by jointly determining the beamforming design at the BS, the MA positions, and the common stream portion for the different users. Due to the non-convex nature of the problem, a two-stage, coarse-to-fine-grained searching algorithm was adopted to produce effective solutions. In \cite{10834523}, the authors examined a short packet Ultra-Reliable and Low-Latency Communication (URLLC) system that leveraged RSMA as a multiple access technique. To overcome the high cost of deploying a large number of antennas at the BS and to produce a more flexible beamforming design, MA technology is deployed at the BS. With the aim of maximizing the sum rate, they presented an optimization problem by jointly optimizing the MA positions along with the MA positions while respecting the URLLC requirements. Then, an effective AO strategy is illustrated to handle the high coupling between the different optimization variables. Utilizing RSMA technology, the work in \cite{amhaz2025enhancingcomprsmaperformancemovable} investigated a DL coordinated multi-point scenario where a set of BSs are jointly serving a set of users acquiring MA technology. The authors formulated an optimization problem with the objective of maximizing the achievable sum rate by determining the optimal beamforming vectors at the cooperating BSs, the MA positions, and the common stream portions while satisfying the quality of service (QoS) requirements. Due to the non-convex nature of the problem stemming from the high interdependence between the variables, a meta-learning-based algorithm that operates without pre-training was deployed in the solution approach.
 
\subsection{Contributions}
To the best of our knowledge, no prior study has investigated the integration of FD-BS-RSMA with MA technology. In particular, we leverage MA technology to examine its effectiveness in mitigating SI and intra-cell interference in FD-BS-RSMA system. Particularly, in a traditional FD-BS system with FPA, residual SI and intra-cell interference primarily handled by either designing beamforming and power allocation \cite{10315060} or by utilizing effective access technique such as RSMA \cite{allu2024robust}. However, utilizing MA with FD-BS along with RSMA provides another degree of freedom to reduce the residual SI impact and intra-cell interference and thus maximizing sum rate. 
In this context, we analyze three distinct scenarios that incorporate MAs in an FD-BS-RSMA system. In scenario 1, multiple UEs, each equipped with a single MA, are served by an FD BS equipped with multiple transmit and receive FPAs. In the second scenario, we consider a system consisting of multiple UEs, each equipped with a single FPA, while the FD BS is equipped with multiple transmit and receive MAs. Finally, in the third scenario,  we consider both the BS and UEs are equipped with MAs, where the BS is equipped with multiple transmit and receive MAs, and each UE has single MA. For all scenarios, we conduct a comprehensive analysis of MA's role in suppressing SI and intra-cell interference, as well as its impact on total sum rate, DL sum rate, and UL sum rate. To the best of our knowledge, this is the first study in the literature to provide a detail investigation of MA technology’s effectiveness in managing interference in FD system.

In this regard, the key contributions of this paper can be
summarized as follows:
\begin{itemize}
\item The system model consists of a FD BS and multiple DL and UL UEs which are served by the BS simultaneously. To cover all the scenarios and to analyze and observe the effect of MAs, as mentioned earlier we consider three scenarios. Then, we formulated three optimization problem covering three scenarios with the objective of maximizing achievable sum rate of all UEs by optimizing transmitting and receiving beamforming vectors at BS, UE transmission power, MA positions and common stream split portions of RSMA while satisfying the minimum data rate constraints at all UEs, common stream rate constraint of RSMA, power budget of BS and UL UEs and inter-MA distance at BS.  
\item The formulated optimization problem is highly non-convex in nature and difficult to tackle using conventional optimization methods due to the high coupling among the different optimization variables. Hence, we propose a gradient-based meta-learning (GML) approach which can handle the optimization problem in a discrete manner by optimizing each variable in a different neural network and having the flexibility of gradient sharing among them. It should be noted that GML has already proven its efficacy in solving large-scale non-convex optimization problems 
\cite{zhu2024robust}. 
\item To evaluate the performance of the model, we perform extensive simulations. First, the convergence of the proposed algorithm is demonstrated, confirming its effectiveness in maximizing the achievable total sum rate while ensuring compliance with the given constraints. Next, the GML-based results are compared with the optimal solution, demonstrating their accuracy and effectiveness in achieving near-optimal performance. 
\end{itemize}
\par The proposed model was analyzed by varying key network parameters, including the BS power budget, UE transmission power, minimum data rate requirements and the residual SI. The results highlighted the effectiveness of MAs in improving SE by effectively handling SI and intra-cell interference and combating fading. Additionally, the results indicate that deploying MAs at the BS yields more favorable outcomes compared to placing them solely at the UE side, as FD communication is more significantly affected by residual SI. However, the optimal performance in terms of SE is achieved when MAs are implemented on both the BS and UE sides. This configuration maximizes the inherent advantages of MA technology by enabling dynamic antenna positioning in regions with lower interference, thereby further enhancing system performance.
\subsection{Paper organization}
This paper is structured as follows: Section II introduces the system model, detailing the rate analysis and the adopted channel model for MA technology. Section III formulates the optimization problem, while Section IV describes the proposed GML algorithm, explaining its key components. Subsequently, Section V presents the simulation results and the performance evaluation. Finally, Section VI concludes the paper with a summary of key findings.
\begin{figure*}
\centering
\includegraphics[width=18cm,height=10cm]{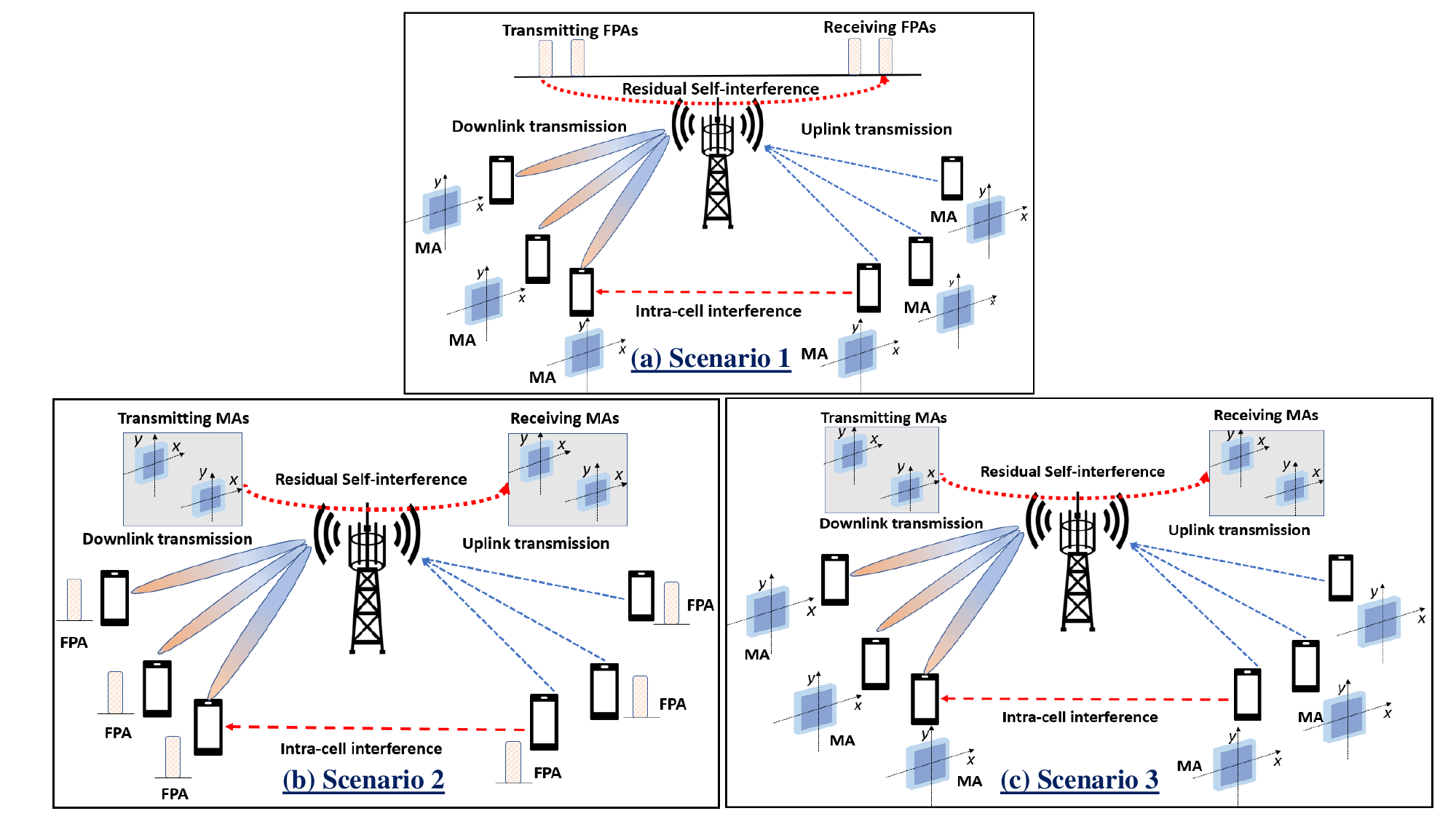}
  \caption{Proposed System Model}
  \label{figs}
\end{figure*}
\section{System Model}
As illustrated in Fig.\ref{figs}, we consider a multi-user FD networking system comprising $D$
DL and $U$
UL UEs. For notational convenience, we define $\mathcal{D}=[1,2,3,...,D]$, $\mathcal{U}=[1,2,3,4,...,U]$, as the sets of DL and UL UEs, respectively.  The BS is equipped with 
$N_T$
  transmit antennas and 
$N_R$
  receive antennas, where
$D \le N_T$ and $U \le N_R$. Additionally, each UE, whether operating in DL or UL, is assumed to be equipped with a single antenna.

As depicted in Fig.\ref{figs}, to assess the impact of MAs in an FD-BS system, we analyze the following three scenarios:
\begin{itemize}
\item \textbf{Scenario 1}: As shown in Fig.\ref{figs}(a), in scenario 1, the BS is equipped with FPAs, while each UE is equipped with a single MA.
\item \textbf{Scenario 2}: Fig.\ref{figs}(b) shows a BS is equipped with 
$N_T$
  transmit MAs and 
$N_R$
  receive MAs. Each UE, whether operating in the DL or UL, is equipped with a single FPA.
\item \textbf{Scenario 3}: Fig.\ref{figs}(c) presents the third scenario where both BS and UEs are equipped with MAs. BS is equipped with $N_T$ transmit and $N_R$ receive MAs, and each UE is equipped with a single MA. 
\end{itemize}
On both sides, MAs are connected to a flexible RF chain via flexible wires such as coaxial cables. Thus, the positions of MAs can be mechanically adjusted with the aid of drive components such as micromotors. A Cartesian coordinate system can be formed to investigate the positions of MAs in the transmit and receive regions. Particularly, the transmit regions and receive regions for MAs of BS can be denoted as $\mathcal{C}_t^{BS}$ and $\mathcal{C}_r^{BS}$, respectively. The coordinates of the transmit MAs of BS are denoted as $\textbf{t}^{BS}_{n_t}=[x_t,y_t]^T \in \mathcal{C}_t^{BS} (1 \le n_t\le N_T)$ and the receive MAs of BS are denoted as $\textbf{r}^{BS}_{n_r}=[x_r,y_r]^T \in \mathcal{C}_r^{BS}  (1 \le n_r\le N_R)$, respectively.  On the other hand, the transmit region for MAs of UL UEs can be denoted by $\mathcal{C}_t^{UL}$ where MA coordinates are represented as $\textbf{t}^{UL}=[x_t,y_t]^T \in \mathcal{C}^{UL}_t$. Meanwhile, the receive region for MAs of DL UEs can be presented by $\mathcal{C}_r^{DL}$ where $\textbf{r}^{DL}=[x_r,y_r]^T \in \mathcal{C}_r^{DL}$. 
%\vspace{-10pt}
\subsection{Transmission Model}
Following the principle of 1-layer RSMA, the BS splits each UE-$d$'s, 
$\forall d \in \mathcal{D}$ message into a common part $A_{c,d}$ and a private part $A_{p,d}$. The common part of all UEs $A_{c,1}, A_{c,2},\dots, A_{c,D}$ is combined together to form a combined common stream $A_c$.  Then it is encoded into a common stream $s_c$ using a codebook that is shared by all DL UEs. It should be noted that $s_c$ is decoded by all DL UEs since it contains parts of messages of all DL UEs. Meanwhile, the private part $A_{p,d}$ of UE-$d$ is independently encoded into a private stream denoted as $s_{p,d}$, $\forall d \in \mathcal{D}$, which is only decoded by UE-$d$.
Specifically, the signal transmitted from BS to the DL UEs is given by,
\begin{equation}
\hat{\textbf{x}}=\textbf{W}_cs_c+\sum_{d \in \mathcal{D}}\textbf{W}_{p,d}s_{p,d}, 
\end{equation}
where $\textbf{W}_c \in \mathbb{C}^{N_T \times 1}$ and $\textbf{W}_{p,d} \in \mathbb{C}^{N_T \times 1}$ represent the transmitting beamforming vectors to transmit a common stream for all DL UEs and a private stream for DL UE-$d$, respectively. 
We assume a FD BS system can transmit and receive signals at the same time. In other words, in the same resource block, the BS transmits a signal to the DL UEs and receives a signal from UL UEs. In particular, utilizing the UL RSMA principle, the messages of UL UEs are split into two sub-messages except for one UL UE. More specifically, we split the messages of $U-1$ UEs into two sub-messages, and the last UE's message is kept without splitting.  Because it has been shown in \cite{liu2020rate} that for a RSMA system with $U$ UL UEs, splitting the messages $U-1$
 is sufficient to achieve the capacity region. Hence, we split each UL UE's message into two sub-messages; the set of sub-messages is denoted as $J \in [1,2]$. Therefore, the transmitted signal of UE-$u, 
 \forall u \in \mathcal{U}$ to the BS can be given by,
\begin{equation}
   x_u= \sum_{j \in J}\sqrt{P_{u,j}}s_{u,j}, \quad \mathrm{if} \quad \{u= 1,2, \cdot \cdot \cdot, U-1\},
\end{equation}
\begin{equation}
    x_u=\sqrt{P_{U}}s_{U}, \quad \mathrm{if} \quad \{u=U\},
\end{equation}
where $P_{u,j}$ is the transmit power for stream $s_{u,j}$ and $P_{U}$ represents the transmit power for stream $s_U$.
From the DL transmission of BS, each DL  UE-$d, \forall d \in \mathcal{D}$ receives the signal. Moreover, UE-$d$ is also impacted by the signals transmitted by all UL UEs.
Hence, the signal received at UE-$d, \forall d \in \mathcal{D}$ can be given as \eqref{eqn4}
\begin{table*}[!ht]
\begin{equation}
%\footnotesize
y_d=\underbrace{\textbf{h}_{d}^H(\textbf{t}_{n_t}^{BS}, \textbf{r}^{DL})(\textbf{W}_cs_c+\textbf{W}_{p,d}s_{p,d})}_{\textrm {desired signal}}+
\underbrace{\sum_{d' \in \mathcal{D}, d' \neq d}\textbf{h}_{d}^H(\textbf{t}_{n_t}^{BS}, \textbf{r}^{DL})\textbf{W}_{p',d'}s_{p',d'}}_{\textrm {interference from private streams of other DL UEs}}
+
\underbrace{\sum_{u \in \mathcal{U}}h_{u,d}(\textbf{t}^{UL},\textbf{r}^{DL})x_u}_{\textrm{intra-cell interference from UL UEs}}+n_d, \label{eqn4}
\end{equation}
\vspace{-.2 in}
\end{table*}
where $\textbf{h}_{d}(\textbf{t}_{n_t}^{BS},\textbf{r}^{DL}) \in \mathbb{C}^{N_T \times 1}$ and  $h_{u,d}(\textbf{r}^{DL}, \textbf{t}^{UL})\in \mathbb{C}^{1\times 1}$ represent the channel coefficient from BS to DL UE-$d$ and UL UE-$u$ to DL UE-$d$, respectively. $n_d \sim \mathcal{CN}(0,\sigma_d^2)$ denotes the additive white Gaussian noise (AWGN) with zero mean and variance $\sigma_d^2$. In the case of Scenario 1, $\textbf{t}_{n_t}^{BS}$ is replaced by a constant location of the antennas in the BS. On the other hand, in the case of Scenario 2, $\textbf{r}^{DL}$, $\textbf{t}^{UL}$ are replaced by the constant locations of the FPAs. For the MA channels for all scenarios, we adopt a field-response-based channel model, which is described in subsection C. 

Alternatively, from the UL transmission of all UE's, the received signal at BS can be denoted by, 
\begin{equation}
%\footnotesize
\hat{\textbf{y}}=\underbrace{\sum_{u \in \mathcal{U}}\textbf{h}_{u}(\textbf{t}^{UL},\textbf{r}_{n_r}^{BS})x_u}_{\textrm{desired signal}}+\underbrace{\textbf{H}^H_{SI}(\textbf{t}_{n_t}^{BS},\textbf{r}_{n_r}^{BS})\hat{\textbf{x}}}_{\textrm{self-interference}}+\textbf{n}_{BS},
\end{equation}
where $\textbf{h}_{u}(\textbf{t}^{UL},\textbf{r}_{n_r}^{BS}) \in \mathbb{C}^{N_R \times 1}$ and $\textbf{H}_{SI}(\textbf{t}_{n_t}^{BS},\textbf{r}_{n_r}^{BS}) \in \mathbb{C}^{N_T \times N_R}$ represent the channel coefficient from UE-$u, \forall u \in \mathcal{U}$ to BS and residual SI channel of BS due to transmitting and receiving at the same time. Meanwhile, there $\textbf{n}_{BS} \sim \mathcal{CN}(0,\sigma_u^2\textbf{I}_{N_R})\in \mathbb{C}^{N_R \times 1}$ is the AWGN with zero mean and variance $\sigma_u^2$. Similar to \eqref{eqn4}, for Scenario 1, $\textbf{t}_{n_t}^{BS}$ and $\textbf{r}_{n_r}^{BS}$ are replaced by the constant location of FPAs. For Scenario 2, $\textbf{t}^{UL}$ is replaced by the constant location of FPAs at UL UEs.
\subsection{Achievable Rate Analysis}
\subsubsection{Achievable rate for DL} The common stream must be decoded by all DL UEs, as it contains segments of messages intended for each DL UE. Hence, each DL UE first decodes the signal of the common stream and then removed it through SIC from the total received signal. Finally,  each DL UE decodes its own private stream. Hence, the achievable rate to decode the common stream at UE-$d, \forall d \in \mathcal{D}$ can be given by \eqref{eqc1}.
\begin{table*}[!ht]
\begin{equation}
%\footnotesize
 R_{c,d}=\log_2\left(1+
\frac{|\textbf{h}_{d}^H(\textbf{t}_{n_t}^{BS}, \textbf{r}^{DL})\textbf{W}_c|^2}{\sum_{d \in \mathcal{D}}|\textbf{h}_{d}^H(\textbf{t}_{n_t}^{BS}, \textbf{r}^{DL})\textbf{W}_{p,d}|^2 +
 \sum_{u=1}^{U-1}(\sum_{j \in J}P_{u,j}|h_{u,d}(\textbf{t}^{UL},\textbf{r}^{DL})|^2)+P_U|h_{U,d}(\textbf{t}^{UL},\textbf{r}^{DL})|^2+\sigma_d^2}\right).\label{eqc1}
  %  \end{split} 
\end{equation}
\end{table*}
After successful decoding of the common stream, it is removed from the total received signal utilizing SIC. The UE-$d$ then decodes its private stream. Hence, the achievable rate to decode the private stream at UE-$d, \forall d \in \mathcal{D}$ is denoted by \eqref{eqp1}.
\begin{table*}[!ht]
\begin{equation}
%\footnotesize
R_{p,d}=\log_2\left(1+\frac{|\textbf{h}_{d}^H(\textbf{t}_{n_t}^{BS}, \textbf{r}^{DL})\textbf{W}_{p,d}|^2}{\sum_{d' \in \mathcal{D}, d' \neq d}|\textbf{h}_{d}^H(\textbf{t}_{n_t}^{BS}, \textbf{r}^{DL})\textbf{W}_{p',d'}|^2+\sum_{u=1}^{U-1}(\sum_{j \in J}P_{u,j}|h_{u,d}(\textbf{t}^{UL},\textbf{r}^{DL})|^2)+P_U|h_{U,d}(\textbf{t}^{UL},\textbf{r}^{DL})|^2+\sigma_d^2}\right).\label{eqp1}
\end{equation}
\end{table*}
In order to guarantee successful decoding of the common stream at all DL UEs, the common stream rate $R_{c,d}$ should be set as the minimum among the common rates of all UEs which can be given as follows,
\begin{equation}
%\footnotesize
    R_{c}=\min\{R_{c,1},\dots,R_{c,D}\}.
\end{equation}
Therefore, we can calculate the total achievable rate for DL-$d, \forall d \in \mathcal{D}$ as follows,
\begin{equation}
%\footnotesize
    R_d^{tot}=C_d+R_{p,d}, \forall d \in \mathcal{D},
\end{equation}
where $C_d$ represents the fraction of $R_c$ assigned to UE-$d$. 
\subsubsection{Achievable rate analysis for UL} It is worth mentioning that in the UL scenario, decoding order plays a crucial role in processing the received signals. Let the decoding order of sub-messages and messages at the BS be represented by the set $\pi = [\pi_{u,j},\pi_{U}|\forall j \in J, \forall u \in \mathcal{U}]$ where the elements are arranged in ascending order. Here, $\pi_{u,j}$ denotes the decoding order of the sub-message $s_{u,j}$, while $\pi_U$ corresponds to the decoding orders of the sub-message $s_U$. 
Specifically, sub-message $s_{u,j}$ is decoded first if its decoding order satisfies $\pi_{u,j} < \pi_{u',j'} $ or $\pi_{u,j} < \pi_{U} $ where $u' \neq u$ and $j \neq j'$, while treating all remaining sub-messages/message as interference. Once $s_{u,j}$ has been successfully decoded, it is removed from the received signal through SIC. The process then continues with the next sub-message or message according to the predefined decoding order. The adopted decoding order for this work is specified in the simulation section. Consequently, the achievable rate to decode sub-message $s_{u,j}$ at BS can be denoted as \eqref{equ1},
\begin{table*}[!ht]
\begin{equation}
%\footnotesize
R_{u,j}=\log_2\left(1+\frac{P_{u,j}|\textbf{Z}_{u,j}^H\textbf{h}_{u}(\textbf{t}^{UL},\textbf{r}_{n_r}^{BS})|^2}{_{j' \neq j}P_{u,j'}|\textbf{Z}_{u,j}^H\textbf{h}_{u}(\textbf{t}^{UL},\textbf{r}_{nr}^{BS})
|^2+I+
\sigma_u^2||\textbf{Z}_{u,j}^H||_2^2}\right). \label{equ1}
 \end{equation}
 \end{table*}
where $\textbf{Z}_{u,j} \in \mathbb{C}^{N_R \times 1}$ represents the receiving beamforming vector at BS to decode sub-message $s_{u,j}$ and 
\begin{equation}
%\footnotesize
\begin{split}
   &I=\sum_{u' \neq u}^{u' \in (U-1)}(\sum_{j \in J}P_{u',j}|\textbf{Z}_{u,j}^H\textbf{h}_{u'}(\textbf{t}^{UL},\textbf{r}_{nr}^{BS})
|^2)+\\
&P_{U}|\textbf{Z}_{u,j}^H\textbf{h}_{U}(\textbf{t}^{UL},\textbf{r}_{nr}^{BS})
|^2+
|\textbf{Z}_{u,j}^H|^2|\textbf{H}_{SI}^H(\textbf{t}_{n_t}^{BS},\textbf{r}_{n_r}^{BS}) 
(\textbf{W}_c+\sum_{d \in \mathcal{D}}\textbf{W}_{p,d})|^2. \notag
\end{split}
\end{equation}In a similar way, the message $s_U$ which is transmitted from UE-$U$ can be decoded at BS which is given as \eqref{equ2}, where $\textbf{Z}_{U} \in \mathbb{C}^{N_R \times 1}$ represents the receiving beamforming vector at BS to decode the sub-message, $s_U$ and $\sigma_U^2$ is the AWGN.
\begin{table*}[!ht]
%\footnotesize
\begin{equation}
R_{U}=\log_2\left(1+\frac{P_{U}|\textbf{Z}_{U}^H\textbf{h}_{U}(\textbf{t}^{UL},\textbf{r}_{n_r}^{BS})|^2}{\sum_{u' \in \mathcal{U}, u' \neq U}(\sum_{j \in J}P_{u',j}|\textbf{Z}_{U}^H\textbf{h}_{u'}(\textbf{t}^{UL},\textbf{r}_{n_r}^{BS})
|^2)+
|\textbf{Z}_{U}^H|^2|\textbf{H}_{SI}^H(\textbf{t}_{n_t}^{BS},\textbf{r}_{n_r}^{BS})
(\textbf{W}_c+\sum_{d \in \mathcal{D}}\textbf{W}_{p,d})|^2+
\sigma_U^2||\textbf{Z}_{U}^H||^2_2}\right).\label{equ2}
\end{equation}
\vspace{-.2 in }
\end{table*}
The total achievable rate for UE-$u$ is given by the sum of its two sub-messages, as follows,
\begin{equation}
%\footnotesize
    R_u=\sum_{j \in J}R_{u,j},  \quad \mathrm{if} \quad \{u= 1,2, \cdot \cdot \cdot, U-1\}. 
\end{equation}
%\vspace{-.2 in }
\subsection{Field-Response Based Channel Model}
Following \cite{ding2024movable}, we model all the communication channels based on the field-response channel model. For MA systems, the channel response is influenced by both the propagation environment and the positioning of the antennas. Similar to \cite{ma2023mimo}, we assume that the antenna's movement region is significantly smaller than the propagation distance between the transmitter and receiver, ensuring that the far-field condition holds at both ends. This assumption is reasonable
because the typical size of the antenna moving region is in the
order of several to tens of wavelengths. As a result, the plane-wave model can be applied to characterize the field response between the transmit and receive regions. This implies that the angles of departure (AoD)s, angles of arrival (AoA)s, and the amplitudes of the complex coefficients for multiple channel paths remain unchanged regardless of the MAs' positions. In contrast, only the phases of the multi-path channels vary within the transmit and receive regions. 
\subsubsection{DL Channels} At the BS side, we denote the number of transmit paths as $L^{BS}_{t}$ from the BS. We assume that the elevation and azimuth AoDs for the $j$-th transmit path between BS and UE-$d$ can be denoted as $\theta_{t,j}^{BS} \in [-\pi/2,\pi/2]$ and $\phi_{t,j}^{BS} \in [-\pi/2,\pi/2], 1 \le j \le L_t^{DL}$. At the UE side, we denote the number of receive paths as $L_r^{d}$. The elevation and azimuth AoAs of the $i$-th receive paths are respectively denoted as $\theta_{r,i}^d \in [-\pi/2,\pi/2]$, $\phi_{r,i}^d \in [-\pi/2,\pi/2]$, $1 \le i \le L_r^{d}$. We define the path-response matrix (PRM) $\boldsymbol{\Sigma}_d \in L_r^d \times L^{BS}_t$ to represent the response from the transmit and reference position $\textbf{t}_{n_t}^{BS}(0)=[0,0]$ to the receive reference position $\textbf{r}^{DL} (0)=[0,0]$. 
Now, the difference between position $\textbf{r}^{DL}=[x_r,y_r]$ and the reference point $\textbf{r}^{DL}[0]$ for the $i$-th receive path, $1 \le i \le L_r^d$ is given as follows:
\begin{equation}
    \rho_{r,i}(\textbf{r}^{DL})=x_rcos\theta_{r,i}sin\phi_{r,i}+y_rsin\theta_{r,i}.
\end{equation}
The channel response of the $i$-th receive path at position $\textbf{r}^{DL}$ exhibits a phase difference of 
$2\pi\rho_{r,i}(\textbf{r}^{DL})/\lambda$ relative to the reference point 
$\textbf{r}^{DL}[0]$, here $\lambda$ refers to the wavelength. To incorporate these phase differences across all 
$L_r^d$
  receive paths, the field-response vector (FRV) within the receive region at DL UE-$d$ is defined as follows:
  \begin{equation}
  %\footnotesize
\textbf{f}_d({\textbf{r}^{DL}})=[e^{j\frac{2\pi}{\lambda}\rho_{r,1}(\textbf{r}^{DL})},\dots,e^{j\frac{2\pi}{\lambda}\rho_{r,L_r^{d}}(\textbf{r}^{DL})}]^T \in \mathbb{C}^{L_r^d \times 1}.
\end{equation}
Thus, we can model the transmit FRV of BS by $\textbf{g}(\textbf{t}_{nt}^{BS})$ and express it as
\begin{equation}
%\footnotesize
      \textbf{g}(\textbf{t}_{n_t}^{BS})=[e^{j\frac{2\pi}{\lambda}\rho_{t,1}(\textbf{t}_{n_t}^{BS})},\dots,e^{j\frac{2\pi}{\lambda}\rho_{t,L^{BS}_{t}}(\textbf{t}_{n_t}^{BS}
    )}]^T \in \mathbb{C}^{{L_t^{BS}}\times 1},
\end{equation}
where $\rho_{t,j}(\textbf{t}^{BS}_{n_t})=x_{t}cos\theta_{t,j}^{BS}sin\phi_{t,j}^{BS}+y_{t}sin\theta_{t,j}^{BS}$, $1\le j \le L_t^{BS}$.
Consequently, the channel vector between the BS and UE-$d$ in DL can be modeled as,
\begin{equation}
%\footnotesize
    \textbf{h}_{d}(\textbf{t}_{n_t}^{BS},\textbf{r}^{DL})= (\textbf{f}_d^H(\textbf{r}^{DL})\boldsymbol{\Sigma}_d\textbf{G}(\textbf{t}_{n_t}^{BS}))^T,
\end{equation}
where $\textbf{G}(\textbf{t}_{n_t}^{BS})=[\textbf{g}(\textbf{t}_1^{BS}),\dots,\textbf{g}(\textbf{t}_{N_T}^{BS})]\in \mathbb{C}^{L_{t}^{BS}\times N_T}$ represents the
transmit field response matrix (FRM) at the BS.
\subsubsection{UL channels} Similarly, for the UL communication, we denote the
number of transmit paths as $L_t^{u}$
from UE. We assume that the elevation and azimuth AoDs for the $j$-th transmit path of UE-$u$ can be denoted as $\theta_{t,j}^{u} \in [-\pi/2,\pi/2]$ and $\phi_{t,j}^{u} \in [-\pi/2,\pi/2], 1 \le j \le L_t^{u}$. At the BS side, we denote the number of receive paths as $L_r^{BS}$. The elevation and azimuth AoAs of the $i$-th receive paths are respectively denoted as $\theta_{r,i}^{BS} \in [-\pi/2,\pi/2]$, $\phi_{r,i}^{BS} \in [-\pi/2,\pi/2]$, $1 \le i \le L_r^{BS}$.  Therefore, we can model the channel between UE-$u$ to BS as follows,
\begin{equation}
\textbf{h}_{u}(\textbf{t}^{UL},\textbf{r}_{n_r}^{BS})=\textbf{F}^H_{b}(\textbf{r}_{n_r}^{BS})\boldsymbol{\Sigma}_{u}\textbf{g}_u(\textbf{t}^{UL}),
\end{equation}
where $\textbf{F}_b=[\textbf{f}_{b}(\textbf{r}_1^{BS}),\dots,\textbf{f}_{b}^{BS}(\textbf{r}_{N_R}^{BS})]\in \mathbb{C}^{L_{r}^{BS}\times N_R}$ is the receive FRM at the BS and receive FRV is 
\begin{equation}
%\footnotesize
\textbf{f}_b(\textbf{r}_{n_r}^{BS})=[e^{j\frac{2\pi}{\lambda}\rho_{r,1}(\textbf{r}_{n_r}^{BS}
    )},\dots,e^{j\frac{2\pi}{\lambda}\rho_{r,L^{BS}_{r}}(\textbf{r}_{n_r}^{BS}
    )}]^T \in \mathbb{C}^{{L}_r^{BS} \times 1}
    \end{equation}and $\rho_{r,i}(\textbf{r}^{DL})=x_{t}cos\theta_{r,i}^{u}sin\phi_{r,i}^{u}+y_{t}sin\theta_{r,i}^{u}, 1 
    \le i \le L_r^{BS}.$
    Meanwhile, transmit FRV can be denoted as 
    \begin{equation}
   % \footnotesize
    \textbf{g}_{u}(\textbf{t}^{UL})=[e^{j\frac{2\pi}{\lambda}\rho_{t,1}(\textbf{t}^{UL}
    )},\dots,e^{j\frac{2\pi}{\lambda}\rho_{t,L^{d}_{t}}(\textbf{t}^{UL}
    )}]^T \in \mathbb{C}^{L_{t}^{d} \times 1}
    \end{equation}, and PRM, $\boldsymbol{\Sigma}_u\in \mathbb{C}^{{L}_r^{BS} \times L_t^d}$ and $\rho_{t,j}(\textbf{t}^{UL})=x_{t}cos\theta_{t,j}^{u}sin\phi_{t,j}^{u}+y_{t}sin\theta_{t,j}^{u}, 1 
    \le j \le L_t^d.$
\subsubsection{SI channel}
    As we consider an FD BS, it suffers from an SI channel. Considering $L_t^{SI}$ the number of transmit paths and $L_r^{SI}$ the number of receive paths of the SI channel, we can model the SI channel as follows,
    \begin{equation}
    %\footnotesize
\textbf{H}_{SI}(\textbf{t}_{nt}^{BS},\textbf{r}_{nr}^{BS})=(\textbf{F}_{SI}^H(\textbf{r}_{nr}^{BS})\boldsymbol{\Sigma}_{SI}\textbf{G}_{SI}(\textbf{t}_{nt}^{BS}))^T,
\end{equation}
where 
\begin{equation}
%\footnotesize
    \textbf{F}_{SI}=[\textbf{f}_{SI}(\textbf{r}_1^{BS}),\dots,\textbf{f}_{SI}(\textbf{r}_{NR}^{BS})] \in \mathbb{C}^{L_r^{SI}\times N_R},
\end{equation}
\begin{equation}
%\footnotesize
\textbf{f}_{SI}(\textbf{r}_{nr}^{BS})=[e^{j\frac{2\pi}{\lambda}\rho_{r,1}(\textbf{r}_{n_r}^{BS}
    )},\dots,e^{j\frac{2\pi}{\lambda}\rho_{r,L^{SI}_{r}}(\textbf{r}_{n_r}
    )}]^T \in \mathbb{C}^{L_r^{SI}\times 1},
    \end{equation} 
    \begin{equation}
    %\footnotesize
\boldsymbol{\Sigma}_{SI}\in \mathbb{C}^{L_r^{SI}\times L_t^{SI}},
\end{equation} 
\begin{equation}
%\footnotesize
    \textbf{G}_{SI}=[\textbf{g}_{SI}(t_{1}^{BS}),\dots,\textbf{g}_{SI}(t_{N_T}^{BS})] \in \mathbb{C}^{L_t^{SI} \times N_T},
    \end{equation}
    \begin{equation}
   % \footnotesize
\textbf{g}_{SI}(\textbf{t}_{n_t}^{BS})=[e^{j\frac{2\pi}{\lambda}\rho_{n_t,1}(\textbf{t}_{nt}^{BS}
    )},\dots,e^{j\frac{2\pi}{\lambda}\rho_{n_t,L^{SI}_{t}}(\textbf{t}_{n_t}^{BS}
    )}]^T \in \mathbb{C}^{L_{t}^{SI} \times 1}.
    \end{equation}
%\vspace{-10pt}
\subsubsection{Interference channels}
UL transmission of UL UEs to BS interferes with the DL transmission of BS to DL UEs. Hence, we also need to model the interference channels from UL UEs to DL UEs. The interference channel model is provided as follows, 
\begin{equation}
%\footnotesize
{h}_{u,d}(\textbf{t}^{UL},\textbf{r}^{DL})=\textbf{f}_u^H(\textbf{r}^{DL})\boldsymbol{\Sigma}_{u_d}\textbf{g}_{d}(\textbf{t}^{UL}),
\end{equation}
where 
\begin{equation}
%\footnotesize
\textbf{f}_{u}(\textbf{r}^{DL})=[e^{j\frac{2\pi}{\lambda}\rho_{r,1}(\textbf{r}^{DL}
    )},\dots,e^{j\frac{2\pi}{\lambda}\rho_{r,L^{ud}_{r}}(\textbf{r}^{DL}
    )}]^T \in \mathbb{C}^{L_r^{ud}\times 1},
    \end{equation}
    \begin{equation}
\boldsymbol{\Sigma}_{u_d} \in \mathbb{C}^{L_r^{ud} \times L_t^{ud}},
\end{equation}
\begin{equation}
%\footnotesize
\textbf{g}_{u}(\textbf{t}^{UL})=[e^{j\frac{2\pi}{\lambda}\rho_{t,1}(\textbf{t}^{UL}
    )},\dots,e^{j\frac{2\pi}{\lambda}\rho_{t,L^{ud}_{t}}(\textbf{t}^{UL}
    )}]^T \in \mathbb{C}^{L_t^{ud}\times 1}.
    \end{equation}
    $L_t^{ud}$ and $L_r^{ud}$ represent the number of transmit and receive paths for the interference channel.
\section{Problem Formulation}
To achieve the full potential of FD communication, we aim to maximize the sum rate of all UEs (DL and UL) by jointly optimizing transmitting and receiving beamforming vectors at BS, the common stream split portion, UE power allocation, and positions of MAs at both the BS and UE's side subject to power budget constraints of BS and UEs and QoS requirements in terms of data rate at all UEs. Hence, we can formulate the optimization problems for "Scenario 1", "Scenario 2", and "Scenario 3" respectively as follows:
\allowdisplaybreaks
\begin{subequations}
\label{prob:P1}
\begin{flalign}
%\footnotesize
\centering
 &\mathcal{P}_1: \max_{\substack{\textbf{P}, \,\textbf{W},\, \textbf{Z}, \, \textbf{c}, \,\textbf{t}, \, \textbf{r}}} \quad \quad \sum_{d \in \mathcal{D}} R_{d}+\sum_{u \in \mathcal{U}} R_u,\:\label{const1} \\
 &\text{s.t.} \quad \textrm{Tr} (\textbf{W}\textbf{W}^H) \le P_{BS},\label{c10}\\
 & \qquad \sum_{j \in J} P_{u,j} \le P_u^{max}, P_{u,j} \ge 0, {\{u=1,2,\dots,U-1\}}, \label{c1}\\
 & \qquad P_U \le P_U^{max}, P_U \ge 0,\label{c2}\\
 & \qquad R_d \ge R_{th,d},  R_u \ge R_{th,u}, \forall d \in \mathcal{D}, \forall u \in \mathcal{U}, \label{c3}\\
 & \qquad \textbf{t}^{UL} \in \mathcal{C}_t^{UL}, \textbf{r}^{DL} \in \mathcal{C}_r^{DL}, \label{c5}\\
 & \qquad |\textbf{Z}_{u,j}|=1,  \{u=1,2,\dots,U-1\}, \forall j \in J,\label{c6}\\
 & \qquad |\textbf{Z}_{U}|=1,  \{u=U\},\label{c11}\\
 & \qquad \sum_{d \in \mathcal{D}} C_d \le R_{c}, C_d \ge 0, \forall d \in \mathcal{D}, \label{c7}
\end{flalign}
\end{subequations}
\allowdisplaybreaks
\begin{subequations}
\label{prob:P1}
\begin{flalign}
\centering
%\footnotesize
 &\mathcal{P}_2: \max_{\substack{\textbf{P}, \,\textbf{W},\, \textbf{Z}, \, \textbf{c}, \,\textbf{t}, \, \textbf{r}}} \quad \quad \sum_{d \in \mathcal{D}} R_{d}+\sum_{u \in \mathcal{U}} R_u,\:\label{const1} \\
 &\text{s.t.} \qquad \textbf{t}_{n_t}^{BS} \in \mathcal{C}_t^{BS}, \textbf{r}_{n_r}^{BS} \in \mathcal{C}_r^{BS}, \label{c4}\\
 & \qquad ||\textbf{t}_{n_t}^{BS}-\textbf{t}_{\hat{n_t}}^{BS}|| \ge DS, 1 \le n_t \neq \hat{n_t}\le N_T, \label{c8} \\
 & \qquad ||\textbf{r}_{n_r}^{BS}-\textbf{r}_{\hat{n_r}}^{BS}|| \ge DS, 1 \le n_r \neq \hat{n_r}\le N_R \label{c9},\\
 & \qquad \eqref{c10}, \eqref{c1}, \eqref{c2}, \eqref{c3}, \eqref{c6}, \eqref{c11}, \eqref{c7}.\notag
\end{flalign}
\end{subequations}
\allowdisplaybreaks
\begin{subequations}
\label{prob:P1}
\begin{flalign}
%\footnotesize
\centering
 &\mathcal{P}_3: \max_{\substack{\textbf{P}, \,\textbf{W},\, \textbf{Z}, \, \textbf{c}, \,\textbf{t}, \, \textbf{r}}} \quad \quad \sum_{d \in \mathcal{D}} R_{d}+\sum_{u \in \mathcal{U}} R_u,\:\label{const1} \\
 &\text{s.t.} \qquad \eqref{c10}, \eqref{c1}, \eqref{c2}, \eqref{c3}, \eqref{c5},\eqref{c6}, \notag\\
 & \qquad \qquad \eqref{c11}, \eqref{c7},\eqref{c4},\eqref{c8},\eqref{c9}, \notag
\end{flalign}
\end{subequations}
where $\textbf{P}=[P_{u,j},P_U], \forall u \in \mathcal{U}, \forall j \in J$, $\textbf{W}=[\textbf{W}_c,\textbf{W}_{p,d}], \forall d \in \mathcal{D}$, $\textbf{Z}=[\textbf{Z}_{u,j},\textbf{Z}_U], \forall u \in \mathcal{U}, \forall j \in J$, $\textbf{t}=[\textbf{t}_{n_t}^{BS},\textbf{t}^{UL}], \forall n_t \in N_T$, $\textbf{r}=[\textbf{r}_{n_r}^{BS},\textbf{r}^{DL}], \forall n_r \in N_R$.
Here, \eqref{c10} refers to the power budget constraint of BS. \eqref{c1} and \eqref{c2} refer to the power budget constraint of UL UEs. Meanwhile, \eqref{c3} ensures the QoS constraints in terms of data rate requirements in both DL and UL UEs. \eqref{c4} and \eqref{c5} limit the ranges of MA movements. \eqref{c6} and \eqref{c11} associate with the receiving beamforming vectors at the BS. \eqref{c7} represents the constraint for successful decoding of common stream by all DL UEs. Finally, \eqref{c8} and \eqref{c9}  ensure that minimum inter-MA distance $DS$ at the BS for practical implementation.
It should be noted that $\mathcal{P}_1, \mathcal{P}_2, \mathcal{P}_3$ are highly nonconvex optimization problem. Specifically, the objective function, QoS constraints, receiving beamforming constraint, and minimum inter-MA distance constraints make the optimization problem highly intractable. Moreover, the optimization problem results in large-scale optimization problems with an increasing number of variables with increasing UEs. Hence, traditional optimization algorithms will suffer from high computational complexity to solve the proposed optimization problem. Thus, we propose a learning-based approach called gradient-based meta-learning, aka GML, to solve the problem with low complexity.
\par It is noteworthy that the formulated problem $\mathcal{P}_3$ is a generalized form of $\mathcal{P}_1$ and $\mathcal{P}_2$ and it covers all the mentioned constraints. Hence, even though we solved all three optimization problems, we describe the solution approach for  the $\mathcal{P}_3$ aka "Scenario 3" only in the next section for the brevity. 
\section{Proposed Solution: Gradient-based Meta Learning (GML) For Resource Allocation}
%\vspace{-.1 in }
Conventional data-driven meta-learning approaches, such as model-agnostic meta-learning (MAML) \cite{yuan2020transfer}, rely heavily on extensive offline pretraining and frequent online adaptations. However, their high energy consumption and sensitivity to variations in data distribution make them impractical for dynamic, latency-sensitive environments. To address these challenges, we employ a model-driven GML framework, as proposed in \cite{zhu2024robust}, which eliminates the need for pretraining and focuses on trajectory optimization rather than optimizing individual variables. This approach has been demonstrated to be effective for non-convex optimization problems \cite{xia2022metalearning}. The GML algorithm follows a three-layer nested cyclic optimization structure, consisting of epoch, outer, and inner iterations, which are further detailed below.
\subsubsection{Inner iterations} Inner iteration is responsible for target variable optimization. We have utilized five separate neural networks to optimize our six target variables. The first neural network is called UE transmission power network (UTPON) to optimize the transmission power of UL UEs and hence optimize $\textbf{P}$. The second neural network is responsible for optimizing the beamforming vectors $\textbf{W}$ at BS, and it is called the beamforming vector network (BVN). The third network is called the receiving beamforming network (RBN), which is responsible for optimizing $\textbf{Z}$. The fourth neural network is called a common rate network (CRN), which is responsible for optimizing the common rate portion in each DL UE. Finally, the last neural network is called the movable antenna network (MOAN). As the variables $\textbf{t}$ and $\textbf{r}$ both have similar characteristics and a purpose to serve, we use only MOAN to optimize them and $\textbf{u}=[\textbf{t},\textbf{r}]$. In each of these networks, optimization of these variables $\textbf{P}, \textbf{W}, \textbf{Z}, \textbf{c}, \textbf{u}$ is updated one after another in each inner iteration.  In each network, optimization
of the target variable starts from the initial value, while
the other variables are inherited from their corresponding
network. The target variables are optimized alternatively in
their corresponding network until convergence. Hence, the
update process in the $j$-th outer iteration is formulated as,
\begin{equation}
%\footnotesize
\textbf{P}^*=\textrm{UTPON}(\textbf{P}^{(i,j)},\textbf{W}^{*},\textbf{u}^*,\textbf{c}^*, \textbf{Z}^*),
\end{equation}
\begin{equation}
%\footnotesize
\textbf{W}^*=\textrm{BVN}(\textbf{P}^*,\textbf{W}^{(i,j)},\textbf{u}^*,\textbf{c}^*,\textbf{Z}^*),
\end{equation}
\begin{equation}
%\footnotesize
\textbf{Z}^*=\textrm{RBN}(\textbf{P}^*,\textbf{W}^*,\textbf{u}^*,\textbf{c}^*,\textbf{Z}^{(i,j)}),
\end{equation}
\begin{equation}
%\footnotesize
\textbf{c}^*=\textrm{CRN}(\textbf{P}^*,\textbf{W}^*,\textbf{u}^*,\textbf{c}^{(i,j)},\textbf{Z}^*),
\end{equation}
\begin{equation}
%\footnotesize
\textbf{t}^*,\textbf{r}^*=\textrm{MOAN}(\textbf{P}^*,\textbf{W}^*,\textbf{u}^{(i,j)},\textbf{c}^*,\textbf{Z}^*),
\end{equation}
where $\textbf{P}^{(i,j)}, \textbf{W}^{(i,j)}, \textbf{Z}^{(i,j)}$,$\textbf{c}^{(i,j)}$,$\textbf{u}^{(i,j)}$ represent $\textbf{P}, \textbf{W}, \textbf{Z}, \textbf{c}, \textbf{u}$ in the $i$-th inner iteration of the $j$-th outer iteration. 
Each of the networks above is iterated $N_i$ times in an inner iteration. As the formulated problem is highly non-convex with coupled optimization variables, the meta-learning structure allows us to solve the optimization problem in a tractable discrete manner. The details of each neural network are provided below.
\par \textbf{UE Transmission Power Network (UTPON):} To optimize transmission power allocation vectors $\textbf{P}$ in $i$-th inner iteration and $j$-th outer iteration, we can rewrite the objective function as follows,
\begin{equation}
%\footnotesize
    \mathcal{P}_4: \max \sum_{d \in \mathcal{D}} R_{d} (\hat{\textbf{W}},\hat{\textbf{u}},\hat{\textbf{c}})+\sum_{u \in \mathcal{U}} R_u (\textbf{P}^{(i,j)},\hat{\textbf{u}},\hat{\textbf{Z}}),\:\label{p1} 
\end{equation}
where $\hat{\textbf{W}}, \hat{\textbf{u}},\hat{\textbf{c}},\hat{\textbf{Z}}$ represent either the initialized or updated parameters. The variable $\textbf{P}$ is updated in two steps. First, we compute the sum rate based on \eqref{p1} where we optimize only the target variable and keep the other variables from initializing or updating from the other network. This sum rate is basically the $\mathcal{L}_{rate}^{(i,j)}$ which is stated in \eqref{37}. Then we calculate the initial loss function $\mathcal{L}^{(i,j)}$ based on \eqref{43} (this has been discussed in details in the section below).  We calculate an initial gradient of $\textbf{P}^{(i,j)}$ with respect to the initial loss function we calculated. This initial gradient of $\textbf{P}^{(i,j)}$ is fed into the neural network. Then, neural network outputs another new gradient/step $\Delta \textbf{P}_{(i,j)}$, which is added to $\textbf{P}^{(i,j)}$ to update it for the next iteration. This update can be given as follows,
\begin{equation}
%\footnotesize
{P}_{u,j}^{(i+1,j)}={P}_{u,j}^{(i,j)}+\Delta{P}_{u,j}^{(i,j)},
\end{equation}
\begin{equation}
%\footnotesize
{P}_{U}^{(i+1,j)}={P}_U^{(i,j)}+\Delta{P}^{(i,j)}_U.
\end{equation}
After updating the UE transmission power, if it exceeds the predefined transmission power budget, we enforce the constraint by projecting the UE transmission power back within the allowable limit. This ensures compliance with the power constraints while maintaining the feasibility of the solution. The projection is performed as follows:
\begin{equation}
%\footnotesize
    {{P}_{u,j}^{(i+1,j)}  = }\begin{cases}{{P}_{u,j}^{(i+1,j)}}, & \mbox{{if }} \mbox{\eqref{c1} satisfies}, \\  {X_1}, & \mbox{{if }} \mbox{{otherwise}}, \end{cases} \label{15}
\end{equation}
\begin{equation}
%\footnotesize
    {{P}_{U}^{(i+1,j)}  = }\begin{cases}{{P}_{U}^{(i+1,j)}}, & \mbox{{if }} \mbox{\eqref{c2} satisfies}, \\  {X_2}, & \mbox{{if }} \mbox{{otherwise}}, \end{cases}\label{16}
\end{equation} 
\begin{equation}
%\footnotesize
    X_1=\frac{P_u^{max}}{\sum_{j \in J}P_{u,j}^{(i+1,j)}}P_{u,j}^{(i+1,j)}, \label{17}
\end{equation}
\begin{equation}
%\footnotesize
    X_2=\frac{P_U^{max}}{P_{U}^{(i+1,j)}}P_{U}^{(i+1,j)}. 
    \label{18}
\end{equation}
\textbf{Beamforming Vector Network (BVN):} Similar to \eqref{p1}, we can denote the objective function in its $i$-th inner iteration and $j$-th outer iteration as follows, where we optimize the target variable $\textbf{W}$,
\begin{equation}
%\footnotesize
    \mathcal{P}_5: \max \sum_{d \in \mathcal{D}} R_{d} ({\textbf{W}^{(i,j)}},\hat{\textbf{u}},\hat{\textbf{c}})+\sum_{u \in \mathcal{U}} R_u \hat{(\textbf{P}},\hat{\textbf{u}},\hat{\textbf{Z}}).\:\label{p2}
\end{equation}
Consequently, we
can obtain the computed gradient from BVN, which can later
be added to the $\textbf{W}^{(i,j)}$, which can be given as follows,
\begin{equation}
%\footnotesize
\textbf{W}^{*}=\textbf{W}^{(i,j)}+\Delta\textbf{W}^{(i,j)}.
\end{equation}
Similar to \eqref{c1} and \eqref{c2}, in order to regulate the updated beamforming to satisfy BS power budget constraint, we project the updated beamforming vector within the power budget limit of BS which is given as follows,
\begin{equation}
%\footnotesize
\textbf{W}^{(i+1,j)}=\sqrt{\frac{P_{BS}}{\textrm{Tr}(\textbf{W}^*{\textbf{W}^*}^H)}}\textbf{W}^*. \label{21}
\end{equation}
\textbf{Receiving Beamforming Network (RBN):}
 In RBN, we optimize \textbf{Z},
while keeping other variables from their the  initial states or updated
from the other networks. Hence, the objective and update of
gradient is measured as follows:
\begin{equation}
%\footnotesize
     \mathcal{P}_6: \max \sum_{d \in \mathcal{D}} R_{d} ({\hat{\textbf{W}}},\hat{\textbf{u}},\hat{\textbf{c}})+\sum_{u \in \mathcal{U}} R_u \hat{(\textbf{P}},\hat{\textbf{u}},{\textbf{Z}^{(i,j)}}),\:\label{p3}
 \end{equation}
 \begin{equation}
 %\footnotesize
\textbf{Z}^{(i+1,j)}=\textbf{Z}^{(i,j)}+\Delta\textbf{Z}^{(i,j)}.
\end{equation}
\textbf{Common Rate Network (CRN):}
In CRN, we optimize $\textbf{c}$, while keeping other variables from its initial states or updated from the other networks. Hence, the objective and update of gradient is measured as follows,
\begin{equation}
%\footnotesize
     \mathcal{P}_7: \max \sum_{d \in \mathcal{D}} R_{d} ({\hat{\textbf{W}}},\hat{\textbf{u}},\textbf{c}^{(i,j)})+\sum_{u \in \mathcal{U}} R_u \hat{(\textbf{P}},\hat{\textbf{u}},\hat{\textbf{Z}}),\:\label{p3}
 \end{equation}
 \begin{equation}
\textbf{c}^{(i+1,j)}=\textbf{c}^{(i,j)}+\Delta\textbf{c}^{(i,j)}.
\end{equation}
\textbf{Movable Antenna Network (MOAN):}
 In this network MOAN, we optimize $\textbf{t}$, $\textbf{r}$ and keep other variables from its initial states or updated from the other networks. Hence, we rewrite the objective and the updated gradient as follows:
 \begin{equation}
% \footnotesize
     \mathcal{P}_8: \max \sum_{d \in \mathcal{D}} R_{d} ({\hat{\textbf{W}}},\textbf{u}^{(i,j)},\hat{\textbf{c}})+\sum_{u \in \mathcal{U}} R_u \hat{(\textbf{P}},\textbf{u}^{(i,j)},\hat{\textbf{Z}}).\:\label{p3}
 \end{equation}
Although MOAN shares the similar structure like the other four networks, due to its small range, it is quite a challenge to keep them within the limited moving region while the meta-agent is performing the exploration. Hence, we have designed a customized regulator to ensure that $\textbf{u}$ within the range of $[-A,A]$ where $A$ denotes the MA moving region, expressed as $\Delta \tilde{\textbf{u}}=\gamma \delta(\Delta \textbf{u})$ where $\gamma$ acts as an amplification operator and $\delta(.)$ represents the Tanh function. This design keeps the MA moving region within a restricted range during the inner iteration, allowing the updated region to be expressed as follows.
 \begin{equation}
\textbf{u}^{(i+1,j)}=\textbf{u}^{(i,j)}+\Delta\tilde{\textbf{u}}.
\end{equation}
\subsubsection{Outer iterations} Outer iterations are responsible for accumulating the meta-loss from the inner iterations. It is important to note that the meta-loss function must comprehensively incorporate both the objective of maximizing the system's sum rate and the constraints necessary to ensure feasibility.  Therefore, we can model the meta-loss function as follows:
\begin{equation}
%\footnotesize
\mathcal{L}^j=-\mathcal{L}_{rate}^{j}+\mathcal{L}_{threshold}^{j}+\mathcal{L}_{norm}^{j}+\mathcal{L}_{common}^{j}+\mathcal{L}_{MA-dist}^{j}, \label{43}
\end{equation}
where $\mathcal{L}_{rate}^{j}$ represents the loss of the objective function and it is expressed as the negative of the all UEs sum rates which can be given as follows,
\begin{equation}
%\footnotesize
    \mathcal{L}_{rate}^{j}=  \sum_{d \in \mathcal{D}} R_{d}+\sum_{u \in \mathcal{U}} R_u,\:\label{37} 
\end{equation}
Next, we add the second term of  meta-loss function to guarantee the constraint \eqref{c3} which can be presented as follows,
\begin{equation}
%\footnotesize
\mathcal{L}_{threshold}^{j}=\sum_{d \in \mathcal{D}}\sum_{u \in \mathcal{U}} \rho_1\lambda_1 (\mathcal{K}_d+\mathcal{K}_u),\:\label{37} 
\end{equation}
\begin{equation}
    \mathcal{K}_d=(R_{th,d}-R_d) \le 0, \forall d \in \mathcal{D},
\end{equation}
\begin{equation}
    \mathcal{K}_u=(R_{th,u}-R_u) \le 0, \forall u \in \mathcal{U},
\end{equation}
\begin{equation}
   {\lambda_1  = }\begin{cases}{0}, & \mbox{{if }} \mbox{{$\mathcal{K}_u, \mathcal{K}_d$}}, \\  {1}, & \mbox{{if }} \mbox{{otherwise}}, \end{cases}
\end{equation}
where $\lambda_1(.)$ represents the indicator function and $\rho_1$ indicates the penalty weight.
The third term in the meta-loss function replaces the constraints \eqref{c6} and \eqref{c11} and it can be presented as,
\begin{equation}
%\footnotesize
    \mathcal{L}_{norm}^{j} =\sum_{u \in \mathcal{U}}\sum_{j \in J} \rho_2\lambda_2  (\mathcal{B}_{u,j}+\mathcal{B}_U),
\end{equation}
\begin{equation}
%\footnotesize
    \mathcal{B}_{u,j}=(|\textbf{Z}_{u,j}|-1) \le 0, \mathcal{B}_{U}=(|\textbf{Z}_{U}|-1) \le 0, 
\end{equation}
\begin{equation}
%\footnotesize
    {\lambda_2  = }\begin{cases}{0}, & \mbox{{if }} \mbox{{$\mathcal{B}_{u,j}, \mathcal{B}_U$}}, \\  {1}, & \mbox{{if }} \mbox{{otherwise}}, \end{cases}
\end{equation}
where $\lambda_2(.)$ represents the indicator function and $\rho_2$ indicates the penalty weight. 
The fourth term of the meta-loss function replaces the constraints \eqref{c7} which can be expressed as,
\begin{equation}
%\footnotesize
    \mathcal{L}_{common}^{j} = \rho_3 \lambda_3 (\Omega),
    \Omega=(\sum_{d \in \mathcal{D}}C_d-R_c)\le 0,
\end{equation}
\begin{equation}
%\footnotesize
    {\lambda_3  = }\begin{cases}{0}, & \mbox{{if }} \mbox{{$\Omega$}}, \\  {1}, & \mbox{{if }} \mbox{{otherwise}}, \end{cases}
\end{equation}
where $\lambda_3(.)$ represents the indicator function and $\rho_3$ indicates the penalty weight. 
Similarly, the fifth term of meta-loss function replaces \eqref{c8} and \eqref{c9}
\begin{equation}
%\footnotesize
    \mathcal{L}_{MA-Dist}^{j} =\sum_{n_t \in N_T} \sum_{n_r \in N_R}\rho_4\lambda_4 (\mathcal{A}_{n_t}+\mathcal{A}_{n_r})
\end{equation}
\begin{equation}
%\footnotesize
   \mathcal{A}_{n_t} =DS-||\textbf{t}_{n_t}^{BS}-\textbf{t}_{\hat{n_t}}^{BS}|| \le 0, 1 \le n_t \neq \hat{n_t}\le N_T,
\end{equation}
\begin{equation}
%\footnotesize
     \mathcal{A}_{n_r} =DS-||\textbf{r}_{n_r}^{BS}-\textbf{r}_{\hat{n_r}}^{BS}|| \le 0, 1 \le n_t \neq \hat{n_t}\le N_R,
\end{equation}
\begin{equation}
%\footnotesize
    {\lambda_4  = }\begin{cases}{0}, & \mbox{{if }} \mbox{{$\Omega$}}, \\  {1}, & \mbox{{if }} \mbox{{otherwise}}, \end{cases}
\end{equation}
where $\lambda_4(.)$ represents the indicator function and $\rho_4$ indicates the penalty weight.
%\vspace{-0.1 in}
\subsection{Epoch iteration:}
This iteration is responsible for updating the neural network parameters. After completing the designated number of outer iterations, the losses are accumulated and averaged accordingly. 
$ \bar{\mathcal{L}}=\frac{1}{N_o}\sum_{j=1}^{N_o}\mathcal{L}^j.\label{60}$
Then, backward propagation is conducted, and the Adam optimizer is used to update the neural networks embedded in these networks, as depicted below:
\begin{equation}
%\footnotesize
\theta_{\textbf{W}}^*=\theta_{\textbf{W}}+\beta_{\textbf{W}}.\textrm{Adam}(\Delta_{\theta_{\textbf{W}}}\mathcal{\bar{L}},\theta_{\textbf{W}}),
\end{equation}
\begin{equation}
\theta_{\textbf{P}}^*=\theta_{\textbf{P}}+\beta_{\textbf{P}}.\textrm{Adam}(\Delta_{\theta_{\textbf{P}}}\mathcal{\bar{L}},\theta_{\textbf{P}}),\label{62}
\end{equation}
\begin{equation}
%\footnotesize
\theta_{\textbf{Z}}^*=\theta_{\textbf{Z}}+\beta_{\textbf{Z}}.\textrm{Adam}(\Delta_{\theta_{\textbf{Z}}}\mathcal{\bar{L}},\theta_{\textbf{Z}}),
\end{equation}
\begin{equation}
\theta_{\textbf{C}}^*=\theta_{\textbf{C}}+\beta_{\textbf{C}}.\textrm{Adam}(\Delta_{\theta_{\textbf{C}}}\mathcal{\bar{L}},\theta_{\textbf{C}}),\label{64}
\end{equation}
\begin{equation}
%\footnotesize
\theta_{\textbf{u}}^*=\theta_{\textbf{u}}+\beta_{\textbf{u}}.\textrm{Adam}(\Delta_{\theta_{\textbf{u}}}\mathcal{\bar{L}},\theta_{\textbf{u}}),
\label{66}
\end{equation}
where $\beta_{\textbf{W}}$, $\beta_{\textbf{P}}$, $\beta_{\textbf{Z}}$, $\beta_{\textbf{C}}$, and $\beta_{\textbf{u}}$, represent the learning rates of the five networks, respectively. The proposed GML is provided in Algorithm 1. 

\textit{Complexity Analysis}: Following \cite{zhu2024robust}, the complexity of the proposed algorithm can be described as follows. The complexity of the proposed GML algorithm lies in calculating the computations for the achievable rates and five neural networks. The first network, UTPON mainly relies on \eqref{equ1}  and \eqref{equ2}. Considering the dimension of $\textbf{Z}_{u,j}$, $\textbf{Z}_{U}$ is  $N_R \times 1$ and hence the computational complexity of multiplying it with $\textbf{h}_u \in N_R \times 1$ is $\mathcal{O}(N_R)$. This calculation is performed $2(U-1)+1$ times for computing the achievable rate and thus, the
overall complexity becomes $\mathcal{O}(2(U-1)+1)^2N_R)$. As gradient calculation in neural network occurs concurrently, the computational complexity of the neural network is
approximately $\mathcal{O}(2(U-1)+1)$. Hence, the overall complexity of UTPON is $\mathcal{O}(2(U-1)+1)^2N_R)$. Similarly, the computational complexity of BVN and CRN is $\mathcal{O}((D+1)^2N_T)$, and RBN is $\mathcal{O}(2(U-1)+1)^2N_R)$. Meanwhile, the computational complexity of MOAN depends on the computation of \eqref{eqc1}, \eqref{eqp1},  \eqref{equ1}, \eqref{equ2} and also its associated calculation of gradient. Hence, its computational complexity is $\mathcal{O}((D+1)^2N_T)+\mathcal{O}(2(U-1)+1)^2N_R)$. Finally, considering
the inner, outer, and epoch iterations, the final time complexity of the proposed algorithm is $N_eN_oN_i(\mathcal{O}(2(U-1)+1)^2N_R)+\mathcal{O}((D+1)^2N_T)+\mathcal{O}((2(U-1)+1)^2N_R)+\mathcal{O}((D+1)^2N_T)+\mathcal{O}((2(U-1)+1)^2N_R)).$
\begin{algorithm}
%\footnotesize
\caption{Proposed algorithm}\label{alg:3}
 Initialize $\theta_{\textbf{P}},\theta_{\textbf{W}},\theta_{\textbf{Z}},\theta_{\textbf{C}},\theta_{\textbf{u}}, \textbf{P}^{(0,1)}$, $\textbf{W}^{(0,1)},\textbf{Z}^{(0,1)},\textbf{C}^{(0,1)}$, $\textbf{u}^{(0,1)}$, inner iterations, $N_i$, outer iterations, $N_o$, epoch iterations, $N_e$;\\
\For {$e={1,2,...,N_e}$}{
$\bar{\mathcal{L}}=0$,MAX$=0$;\\
\For {$j=1,2,...N_o$}{
  $\textbf{P}^{(0,j)}=\textbf{P}^{(0,1)}$,$\textbf{W}^{(0,j)}=\textbf{W}^{(0,1)}$ ;\\
$\textbf{Z}^{(0,j)}=\textbf{Z}^{(0,1)}$,
$\textbf{C}^{(0,j)}=\textbf{C}^{(0,1)}$;\\
$\textbf{u}^{(0,j)}=\textbf{u}^{(0,1)}$;\\

     \For {$i= 1,2,...,N_i$}{
      %$R_{\textbf{P}}^{(i-1,j)}=\max \sum_{d \in \mathcal{D}} R_{d} (\hat{\textbf{W}},\hat{\textbf{u}},\hat{\textbf{c}})+\sum_{u \in \mathcal{U}} R_u (\textbf{P}^{(i-1,j)},\hat{\textbf{u}},\hat{\textbf{Z}})$;\\
      Solve $\mathcal{P}_4$;\\
      $\Delta \textbf{P}^{(i-1,j)}$= $\textrm{UTPON}$
      $(\Delta_{\textbf{P}}R_{\textbf{P}}^{(i-1,j)})$;\\
      $\textbf{P}^{(i,j)}=\textbf{P}^{(i-1,j)}+\Delta \textbf{P}^{(i-1,j)}$;
      }
      $\textbf{P}^*=\textbf{P}^{(N_i,j)}$;\\
      Apply projection if applicable as \eqref{15}-\eqref{18}\\
      \For {$i= 1,2,...,N_i$}{
      %$R_{\textbf{W}}^{(i-1,j)}= \max \sum_{d \in \mathcal{D}} R_{d} ({\textbf{W}^{(i-1,j)}},\hat{\textbf{u}},\hat{\textbf{c}})+\sum_{u \in \mathcal{U}} R_u \hat{(\textbf{P}},\hat{\textbf{u}},\hat{\textbf{Z}})$;\\
      Solve $\mathcal{P}_5$;\\
      $\Delta \textbf{W}^{(i-1,j)}$= $\textrm{BVN}(\Delta_{\textbf{W}}R_{\textbf{W}}^{(i-1,j)})$;\\
      $\textbf{W}^{(i,j)}=\textbf{W}^{(i-1,j)}+\Delta \textbf{W}^{(i-1,j)}$;
      }
      $\textbf{W}^*=\textbf{W}^{(N_i,j)}$;\\
      Apply projection as \eqref{21}\\
        \For {$i= 1,2,...,N_i$}{
     % $R_{\textbf{Z}}^{(i-1,j)}=\max \sum_{d \in \mathcal{D}} R_{d} ({\hat{\textbf{W}}},\hat{\textbf{u}},\hat{\textbf{c}})+\sum_{u \in \mathcal{U}} R_u \hat{(\textbf{P}},\hat{\textbf{u}},{\textbf{Z}^{(i-1,j)}})$;\\
     Solve $\mathcal{P}_6$;\\
      $\Delta \textbf{Z}^{(i-1,j)}$= $\textrm{RBN}$
      $(\Delta_{\textbf{Z}}R_{\boldsymbol{\Phi}}^{(i-1,j)})$;\\
      $\textbf{Z}^{(i,j)}=\textbf{Z}^{(i-1,j)}+\Delta \textbf{Z}^{(i-1,j)}$;
      }
    $\textbf{Z}^*=\textbf{Z}^{(N_i,j)}$;\\

        \For {$i= 1,2,...,N_i$}{
     % $R_{\textbf{C}}^{(i-1,j)}=\max \sum_{d \in \mathcal{D}} R_{d} ({\hat{\textbf{W}}},\hat{\textbf{u}},\textbf{c}^{(i-1,j)})+\sum_{u \in \mathcal{U}} R_u \hat{(\textbf{P}},\hat{\textbf{u}},\hat{\textbf{Z}}),$;\\
     Solve $\mathcal{P}_7$;\\
      $\Delta \textbf{C}^{(i-1,j)}$= $\textrm{CRN}$
      $(\Delta_{\textbf{}{C}}R_{\textbf{C}}^{(i-1,j)})$;\\
      $\textbf{C}^{(i,j)}=\textbf{C}^{(i-1,j)}+\Delta \textbf{C}^{(i-1,j)}$;
      }
    $\textbf{C}^*=\textbf{C}^{(N_i,j)}$;\\

        \For {$i= 1,2,...,N_i$}{
     % $R_{\textbf{u}}^{(i-1,j)}=\max \sum_{d \in \mathcal{D}} R_{d} ({\hat{\textbf{W}}},\textbf{u}^{(i-1,j)},\hat{\textbf{c}})+\sum_{u \in \mathcal{U}} R_u \hat{(\textbf{P}},\textbf{u}^{(i-1,j)},\hat{\textbf{Z}})$;\\
     Solve $\mathcal{P}_8$;\\
      $\Delta \textbf{u}^{(i-1,j)}$= $\textrm{MOAN}$ $(\Delta_{\textbf{u}}R_{\textbf{u}}^{(i-1,j)})$;\\
      $\textbf{u}^{(i,j)}=\textbf{u}^{(i-1,j)}+\Delta \textbf{u}^{(i-1,j)}$;
      }
    $\textbf{u}^*=\textbf{u}^{(N_i,j)}$;\\
   Calculate $\mathcal{L}^j$ as \eqref{43};\\
   $\bar{\mathcal{L}}=\bar{\mathcal{L}}+\mathcal{L}^j$;\\
   \textbf{If} {$-\mathcal{L}^j>MAX$}\\
        \quad MAX=$-\mathcal{L}^j$;\\
        \quad 
        $\textbf{P}_{opt}=\textbf{P}^*$,
$\textbf{W}_{opt}=\textbf{W}^*$,
$\textbf{Z}_{opt}=\textbf{P}^*$,$\textbf{C}_{opt}=\textbf{C}^*$, $\textbf{u}_{opt}=\textbf{u}^*$;\\
   \textbf{end if}
   }
     Calculate  $\bar{\mathcal{L}}=\frac{1}{N_o}\sum_{j=1}^{N_o}\mathcal{L}^j$;\\
update following \eqref{62}-\eqref{66}, 
}
return $\quad \textbf{P}_{opt},\textbf{W}_{opt},\textbf{Z}_{opt},\textbf{C}_{opt}, \textbf{t}_{opt}, \textbf{u}_{opt}$
\end{algorithm}
%\vspace{-.1 in}
\section{Simulation results}
In this section, extensive simulations are carried out to evaluate the performance of the proposed system. We first provide the simulation setup and then present the numerical results for verifying the efficacy of the proposed scheme. The simulation parameters are summarized in Tables I and II. 

In the simulation, we assume two DL UEs and two UL UEs. The UEs are randomly and uniformly distributed in a cell centered on the FD BS located at (0,0) in an area of $200m \times 200m$. The FD BS is equipped with an equal number of transmit and receive antennas, i.e., $N_T=N_R$. Following \cite{zhu2023modeling}, we adopt the same channel where we assume that  $L_t^{BS}=L_r^{BS}=L_t^{DL}=L_t^u=L_t^d=L_r^{SI}=L_t^{SI} \equiv L$. In this way, every PRM $\boldsymbol{\Sigma}_d$, $\boldsymbol{\Sigma}_u$, $\boldsymbol{\Sigma}_{SI}$, $\boldsymbol{\Sigma}_{u_d}$ becomes a diagonal matrix where $\boldsymbol{\Sigma}_d = \textrm{diag}\{\sigma_{1,1},\sigma_{1,1},\dots,\sigma_{L,L}\}$, $\boldsymbol{\Sigma}_u = \textrm{diag}\{\sigma_{1,1},\sigma_{1,1},\dots,\sigma_{L,L}\}$, $\boldsymbol{\Sigma}_{SI} = \textrm{diag}\{\sigma_{1,1},\sigma_{1,1},\dots,\sigma_{L,L}\}$, $\boldsymbol{\Sigma}_{u_d} = \textrm{diag}\{\sigma_{1,1},\sigma_{1,1},\dots,\sigma_{L,L}\}$. Moreover, each diagonal element follows a Circularly symmetric complex gaussian (CSCG) distribution $\mathcal{CN} (0, \frac{\rho}{L})$. Meanwhile, we model the UL and DL channels as each element of the PRV follows the CSCG distribution $\mathcal{CN} (0, 
 \frac{g_0 d_{d/u}^{-\alpha}}{L})$, where $g_0$ represents the path loss at the reference distance of 1m, $\alpha$ is the path loss exponent, and $d_{u/d}$ represents the propagation distance from BS to DL UE-$d$ or the distance from UL UE-$u$ to BS. Similarly, we can model the channel between each UL UE-$u$ and DL UE-$d$ where each PRV element follows the CSCG distribution $\mathcal{CN} (0, 
 \frac{g_{0_1} d_{ud}^{-\alpha_1}}{L})$. Meanwhile, PRM of SI channel follows the CSCG distribution $\mathcal{CN} (0, 
 \frac{SI}{L})$. We assume that the AoDs and AoAs are to be independent and identically distributed random variables within the interval of $[0,2\pi]$. The numerical results are averaged over 100 independent and random channel realizations. Learning rates for all neural networks are set to 0.001. Following \cite{liu2020rate}, we adopt the optimal decoding order of the two-user case for UL UEs which is given as follows: $s_{1,1} < s_{2} < s_{1,2}$. Unless otherwise specified, the default simulation parameters are provided in Table I and Table II. 
 In our simulation results, we denote our proposed "\textit{Scenario 1}" as "\textbf{FD RSMA-MA UE side}", "\textit{Scenario 2}" as "\textbf{FD RSMA-MA BS side}", and "\textit{Scenario 3}" as "\textbf{FD RSMA-MA both side}". Finally, we compare our three scenarios with the traditional FD-RSMA-FPA scheme, which represents that BS and UEs are equipped with FPAs. It should be noted that all four of these schemes consist of the same number of antennas on the BS and UE sides.
 \begin{table}[!t]
%\footnotesize
\centering
\caption{Simulation parameters}
\begin{tabular}{ |l|c|c|c|}
 \hline
  \textbf{Parameter}& \textbf{Value} &\textbf{Parameter} & \textbf{Value}\\
 \hline
 $N_T$, $N_R$ & 4 & $g_0$, $g_{0_1}$ & -40 dB, -50 dB\\
 \hline
 $\lambda$, $A$ & 0.01, $2\lambda$ & $\alpha$, $\alpha_1$ & 2.8, 3.5\\
 \hline
 $R_{th,d}, R_{th,u}$ & 1 bps/Hz & SI & -90 dB\\
 \hline
  $\sigma_d^2$, $\sigma_u^2$, $\sigma_U^2$ & -90 dBm & $S$, $DS$ & 20, $\lambda/2$\\
 \hline
 $L_r$, $L_t$ & 6 & $P^{max}_u$, $P_{BS}$ & 23 dBm, 30 dBm\\
 \hline
\end{tabular}
\end{table}
\begin{table}
%\footnotesize
\centering
\caption{Number of neurons in the neural networks}
\begin{tabular}{ |l|p{15mm}|c|c|p{15mm}| }
 \hline
  \textbf{Layer} & \textbf{Input} & \textbf{Linear } & \textbf{ReLU} & \textbf{Output} \\
  \hline
UTPON & $2(U-1)+1$ & 200 & 200& $2(U-1)+1$ \\
 \hline
BVN  & $2K$ & 200 & 200 & $2K$ \\
 \hline
 MOAN & $N_T+N_R \newline +D+U$ & 200 & 200 & $N_T+N_R \newline +D+U$\\
 \hline
RBN & $2(U-1)+1$ & 200 & 200 & $2(U-1)+1$\\
 \hline
 CRN & $D$& 200& 200& $D$\\
 \hline
\end{tabular}
\end{table}
 \begin{figure}[!ht]
    \centering
\includegraphics[width=0.9\linewidth]{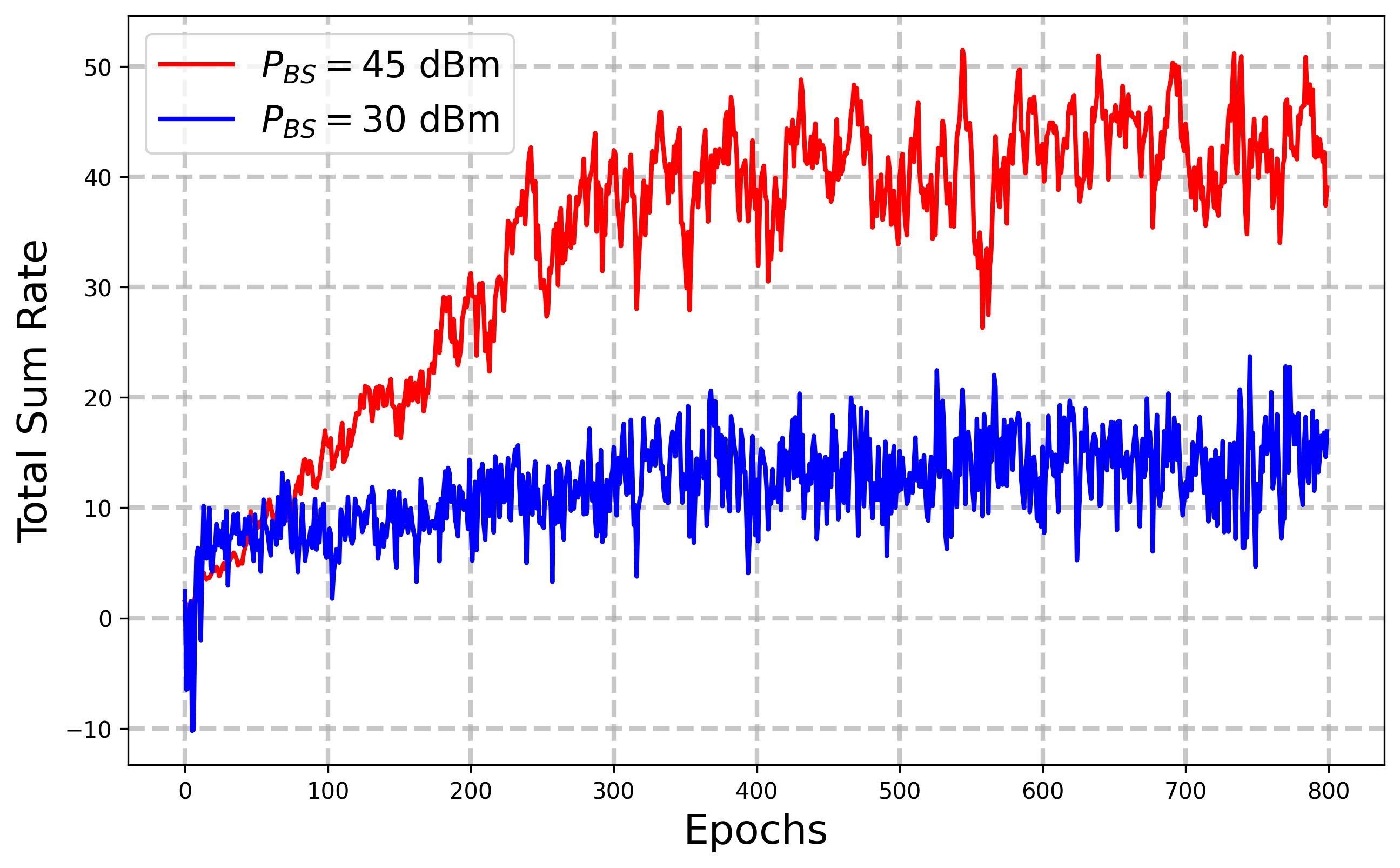}
    \caption{Convergence of the proposed algorithm}
    \label{fig3}
\end{figure}
\begin{figure*}
\centering
\hspace{-0.35cm}\subfigure[] {\centering\includegraphics[width=0.35\textwidth]{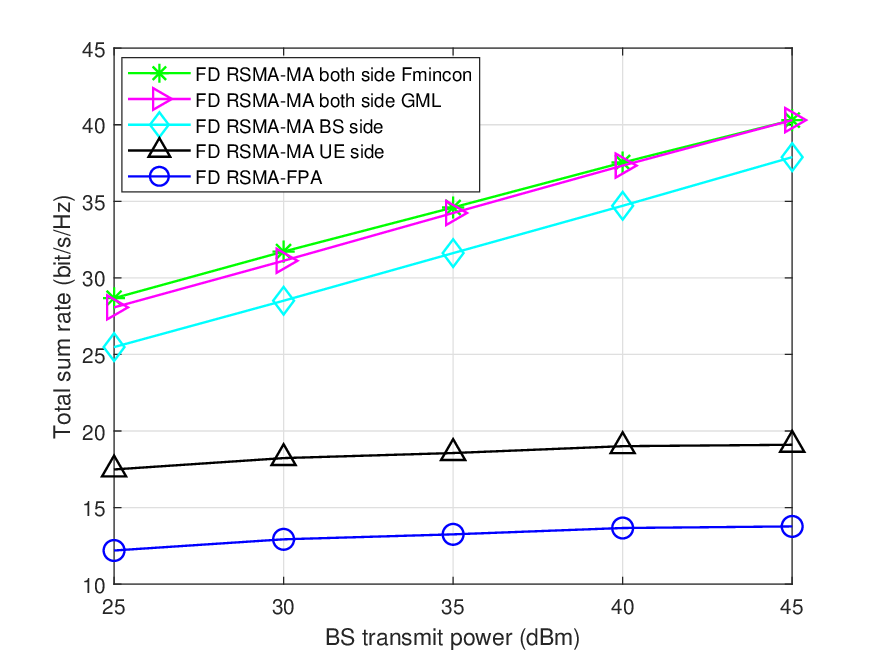}} 
\hspace{-0.5cm}\subfigure[] {\centering\includegraphics[width=0.35\textwidth]{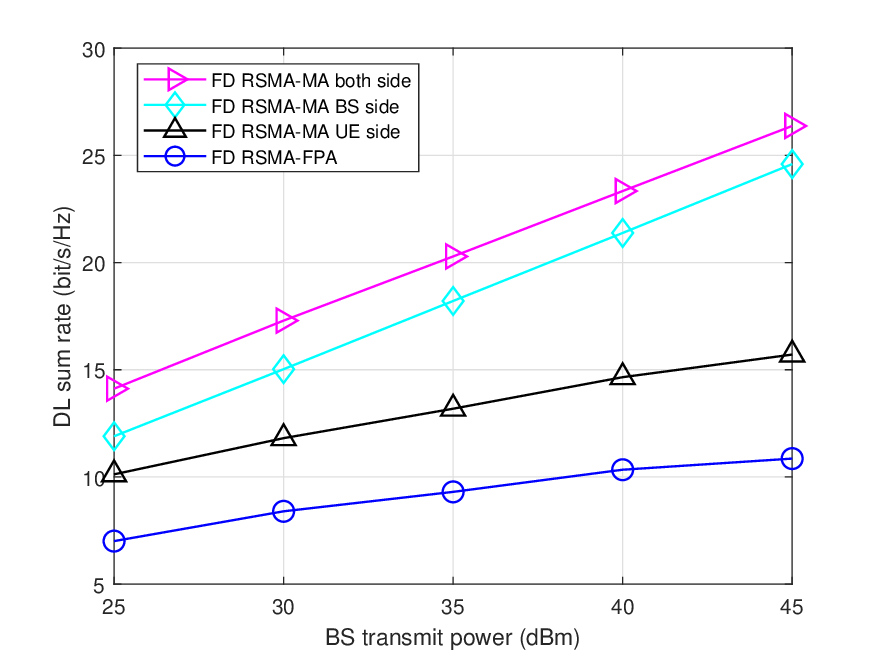}} 
\hspace{-0.5cm}\subfigure[] {\centering\includegraphics[width=0.35\textwidth]{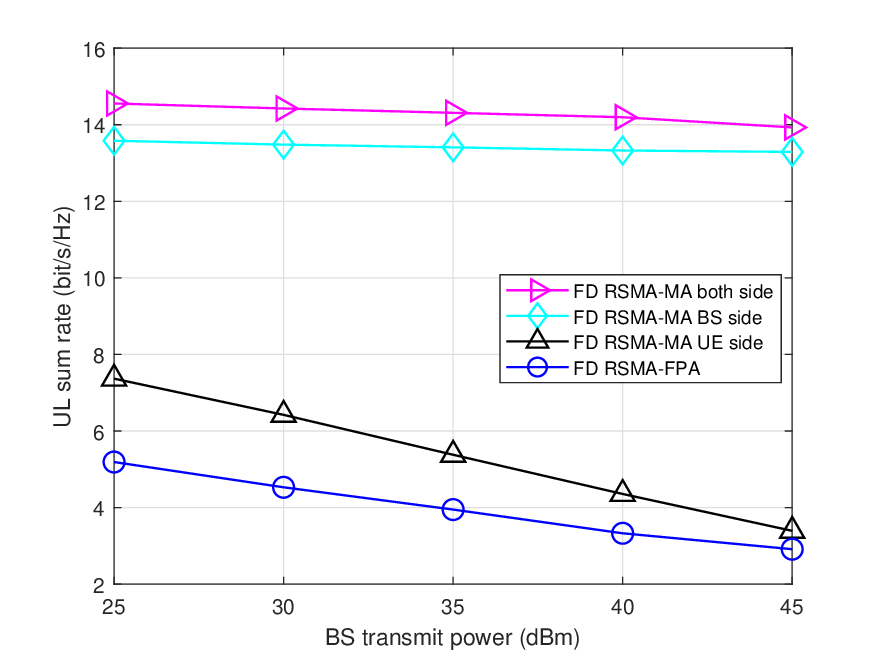}}
\caption{(a) Total achievable sum rate vs BS transmit power (b) DL sum rate vs BS transmit power  (c) UL sum rate vs BS transmit power}
\label{fig7}
\vspace{-.2 in}
\end{figure*}

\begin{figure*}
\centering
\hspace{-0.35cm}\subfigure[] {\centering\includegraphics[width=0.35\textwidth]{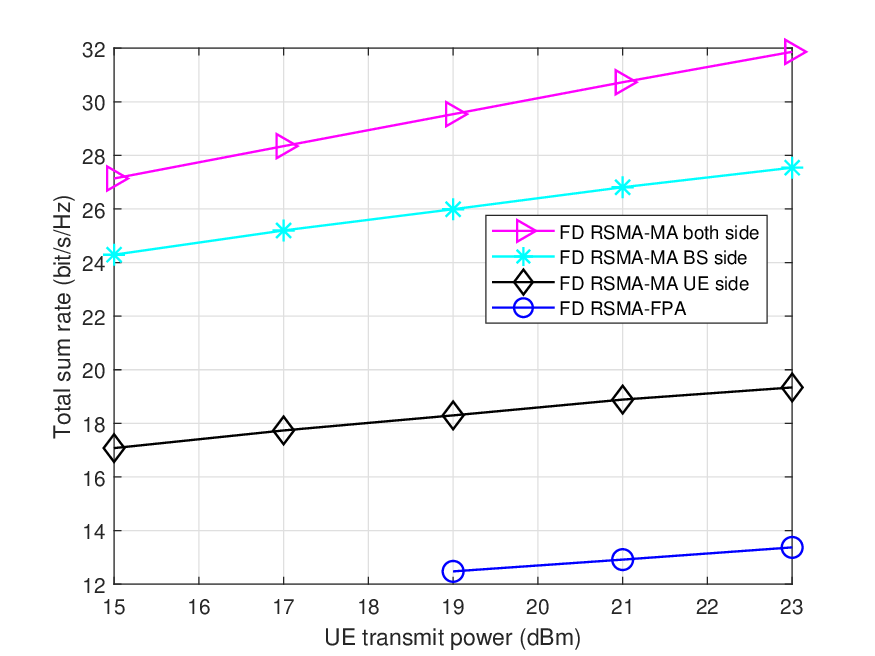}} 
\hspace{-0.5cm}\subfigure[] {\centering\includegraphics[width=0.35\textwidth]{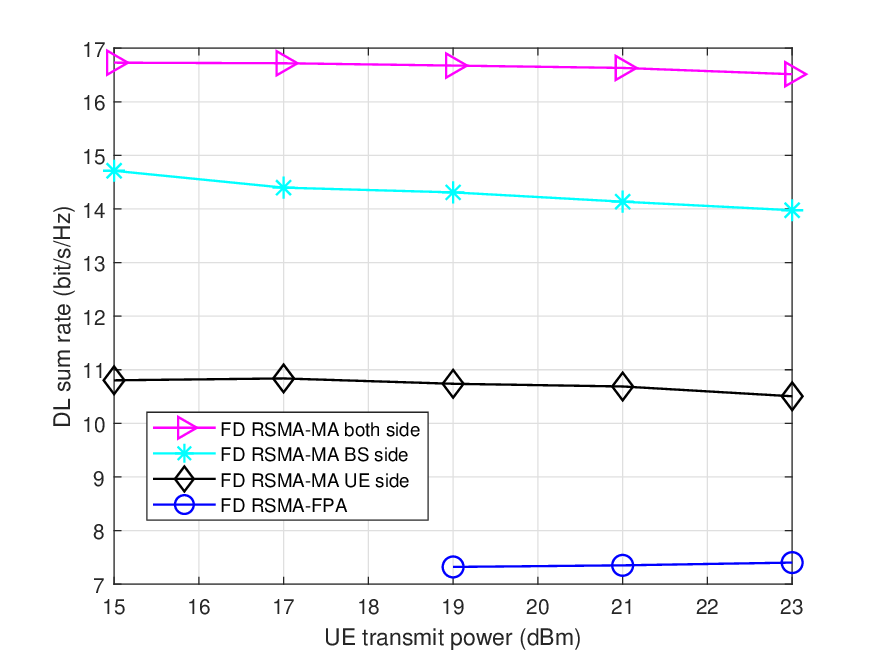}}  
\hspace{-0.5cm}\subfigure[] {\centering\includegraphics[width=0.35\textwidth]{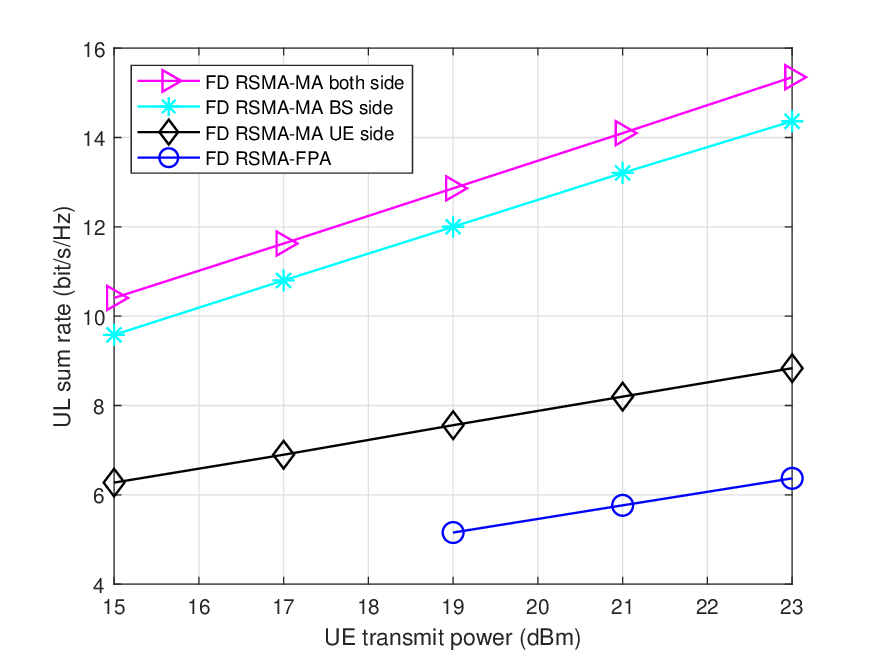}}
\caption{(a) Total achievable sum rate vs UL UE transmit power (b) DL sum rate vs UL UE transmit power (c) UL sum rate vs UL UE transmit power}
\label{fig10}
\vspace{-.2 in}
\end{figure*}
%\vspace{-.1 in}
\subsection{Convergence Analysis}
Fig. \ref{fig3} presents the convergence behavior of our proposed GML algorithm for the proposed FD-BS RSMA with MA. For brevity, we provide only the convergence plot of "Scenario 3" where both BS and UE are equipped with MAs. We plot the convergence of "Scenario 3" for two power budgets of BS: 30 dBm and 45 dBm, respectively. It should be noted that convergence is plotted for a single channel realization only. It can be seen from the figure that when $P_{BS}=30$ dBm, our proposed algorithm converges after 300 epochs. Meanwhile, for $P_{BS}=45$ dBm, the algorithm stops increasing significantly after 500 epochs. Despite the inherent complexity and non-convexity of the formulated optimization problem, our proposed algorithm demonstrates fast convergence, typically within 300 to 500 epochs.
%\vspace{-.1 in }
\subsection{Impact of BS Transmit Power}
Fig. \ref{fig7}(a) illustrates the relationship between the BS transmit power and the total sum rate when the residual SI coefficient is -60 dB. It can be observed that FD RSMA-MA on both sides and FD-RSMA-MA BS side achieve significantly higher sum rates as the BS transmission power increases. This performance gain is attributed to the presence of MAs at the BS, which enhance adaptability in mitigating the negative effects of residual SI in FD communication. By contrast, FD-RSMA-MA UE side and FD-RSMA-FPA, which rely on FPAs at the BS, exhibit limited capability in counteracting the adverse impact of residual SI, thereby constraining their achievable sum rates. In addition to the extraordinary benefits achieved by deploying MA in the FD-BS-RSMA system, our proposed GML scheme achieved near-optimal performance with around 98.5\% accuracy. This conclusion is drawn by comparing the results of the FD-RSMA-BS-MA model on both sides when solving the optimization problem using the GML approach versus the Fmincon solver, which was initialized with 100 starting points to ensure solutions close to the optimal value. The findings are further validated by Fig. \ref{fig7}(b) and \ref{fig7}(c). Specifically, Fig. \ref{fig7}(b) presents the performance of the DL sum rate, revealing that as the BS power budget increases, FD-RSMA-MA both side and the FD-RSMA-MA BS side experience a significant increase in the DL sum rate, while the FD-RSMA-MA UE side and FD-RSMA-FPA exhibit only a moderate increase. 

Similarly, Fig. \ref{fig7}(c) examines the impact of BS transmit power on UL sum rate. It shows that increasing the BS power results in greater SI at the BS receiver due to FD operation, which results in a lower sum rate at UL reception. Notably, FD-RSMA-MA both side and FD-RSMA-MA BS side effectively mitigate this effect through optimal MA placement, ensuring that SI has minimal impact on the UL sum rate, even as BS power increases. In contrast, FD-RSMA-MA UE side and FD-RSMA-FPA, which are equipped with FPAs at BS, suffer a sharp decline in sum rate as BS transmission power increases, further revealing the adaptability of MA in interference management. At 30 dBm of BS power budget, FD-RSMA-MA both side achieves around 9.15\% performance gain over FD-RSMA-MA BS side, 73.94\% gain over FD-RSMA-MA UE side and 145.32\% gain over FD-RSMA-FPA in terms of total achievable sum rate. Meanwhile, at 30 dBm of BS power budget, performance gain of FD-RSMA-MA BS side over FD-RSMA-MA UE side is around 56.67\% and around 120.54\% gain over FD-RSMA-FPA in terms of total sum rate. Finally, FD-RSMA-MA UE side achieves around 41.03\% performance gains over FD-RSMA-FPA in terms of total sum rate at 30 dBm of BS power budget.
%\vspace{-.1 in}
\begin{figure*}
\centering
\hspace{-0.35cm}\subfigure[] {\centering\includegraphics[width=0.35\textwidth]{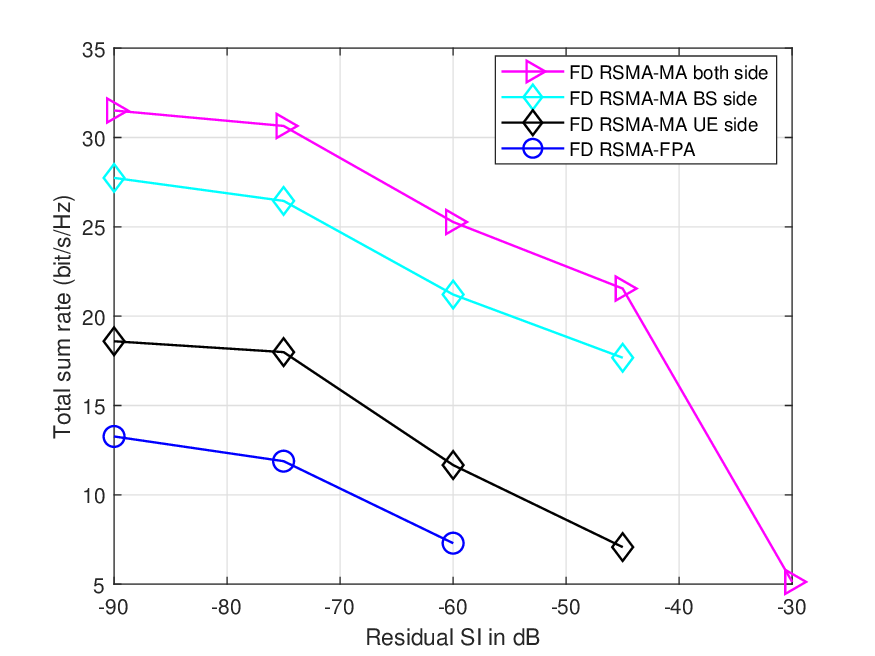}} 
\hspace{-0.5cm}\subfigure[] {\centering\includegraphics[width=0.35\textwidth]{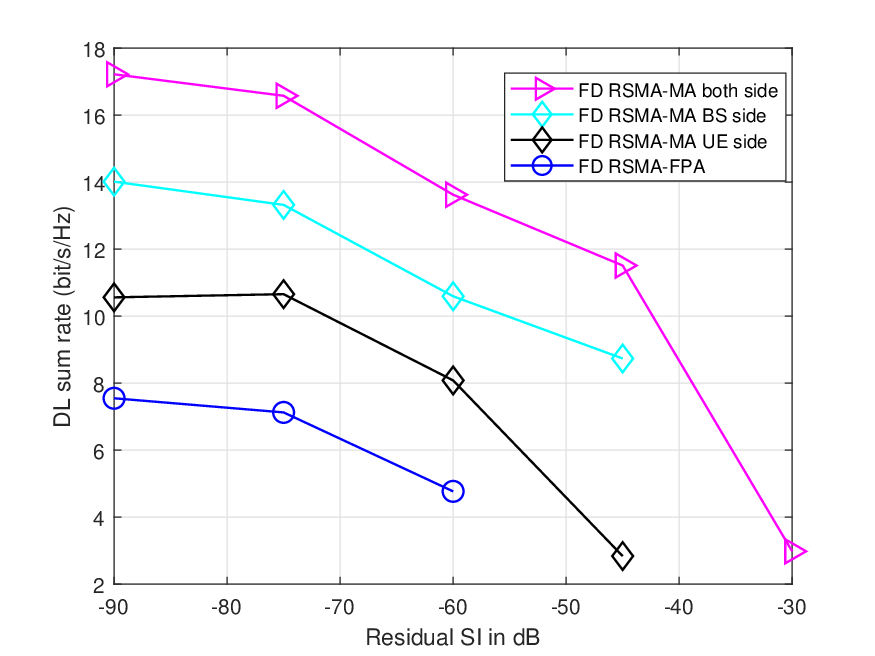}}  
\hspace{-0.5cm}\subfigure[] {\centering\includegraphics[width=0.35\textwidth]{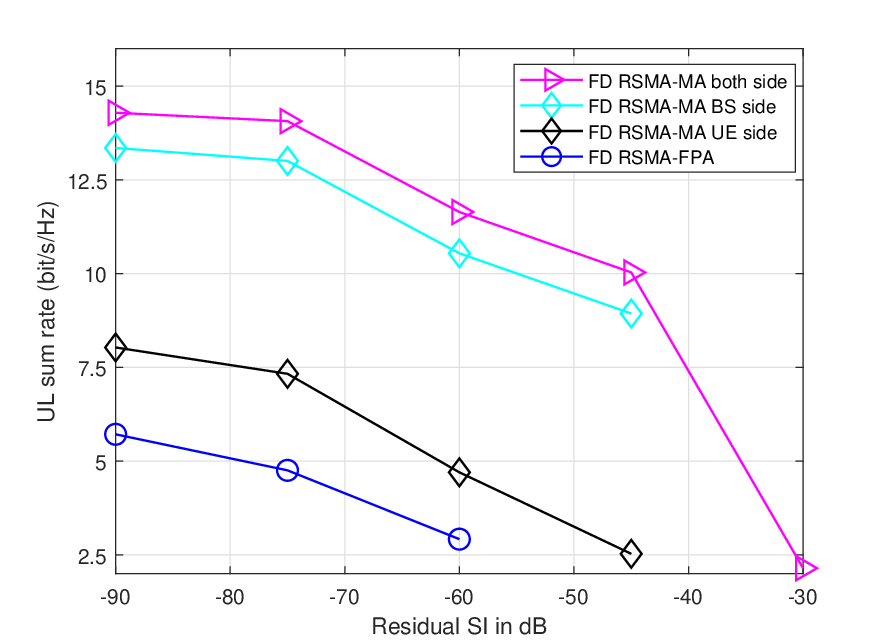}}
\caption{(a) Total achievable sum rate vs Residual SI (b) DL sum rate vs Residual SI (c) UL sum rate vs Residual SI}
\label{fig4}
\vspace{-.2 in}
\end{figure*}
%\vspace{-.2in }
\subsection{Impact of UE Transmission Power}
In Fig. \ref{fig10} we analyze the impact of the UE transmission power on the total sum rate, as well as the individual DL and UL sum rates, by varying the UE transmission power from 15 dBm to 23 dBm. The results indicate that as the UE transmission power increases, the total sum rate also improves. Among all evaluated schemes, FD RSMA-MA both side achieves the highest total sum rate, as it benefits from deploying MA at both the transmitter and receiver sides. This configuration fully exploits spatial diversity, leading to superior performance. In Fig. \ref{fig10}(a), comparing this with the scheme that considers MA only at the BS side, we observe that at lower UE power budgets, the performance gap between these two schemes remains relatively modest. However, as the UE transmission power increases, the performance gap widens. This is because FD-RSMA MA on both sides, which incorporates MA at both ends, effectively mitigates SI and intra-cell interference, whereas FD-RSMA-MA BS side, which contains MA only at the BS, can address SI but struggles with intra-cell interference as the UEs rely solely on FPAs.

Furthermore, in the FD-RSMA-MA UE side, where MA is implemented only at the UE side, intra-cell interference is handled more effectively compared to FD-RSMA-FPA. However, the results confirm that MA deployment on both the transmitter and receiver sides consistently outperforms all other configurations, as it maximizes interference mitigation and spatial diversity gains. At 23 dBm of UE power budget, our proposed FD-RSMA-MA on both side achieves around 15.68\% performance gains over the proposed FD-RSMA-MA BS side, 64.74\% performance gains over the FD-RSMA-MA UE side and 138.29\% performance gains over FD-RSMA FPA in terms of total sum rate. Meanwhile, the proposed FD-RSMA-MA UE side achieves around 44.67\% improvement over FD-RSMA-FPA in terms of sum rate at 23 dBm of UE transmission power budget. It should be noted that for the case of FD RSMA-FPA, when the UE transmission power is around 15 dBm to 19 dBm, it fails to produce to any feasible solutions while meeting the QoS requirements. Meanwhile, Fig. \ref{fig10}(b) presents only DL sum rate vs the UE transmission power. It can be seen from the figure that as we increase the transmission power of UL UEs, the DL sum rate tends to decrease a small amount, even though the BS transmission power remains the same. This is because as we increase the power budget of UEs, intra-cell interference coming from UL UE transmission towards the DL UE reception tends to increase, resulting in a lower sum rate. In this circumstance, FD-RSMA-MA both side and FD-RSMA-MA UE side seem to handle better intra-cell interference as both of them are equipped with MA in comparison to the schemes who have only FPAs at UE side (FD-RSMA-MA BS side and FD-RSMA-FPA).  On the other hand, Fig. \ref{fig10}(c) shows the behavior of UL sum rate when we increase the UE's transmission power. It presents that every scheme increases, with a higher rate of increase exhibited by the two schemes with MAs on the BS side.
\subsection{Impact of Residual SI}
Fig. \ref{fig4} examines the impact of residual SI on the sum rate performance. In Fig. \ref{fig4}(a), we analyze the total sum rate of the system as the residual SI varies from -90 dB to -30 dB. It is evident that as residual SI increases, the sum rate of all schemes declines. This degradation occurs because a higher SI level forces the BS to reduce its transmission power to mitigate its negative effects, ultimately leading to a lower achievable sum rate.
Fig. \ref{fig4}(b) further investigates the effect of SI on the DL sum rate. The results confirm that SI negatively impacts DL transmission in an FD system. As the SI level increases, the DL sum rate decreases due to the inherent FD operation. Although SI is measured at the UL reception side, its effect indirectly influence the DL transmission. When the residual SI at the BS receiver is high, the BS reduces its transmission power to limit SI-induced degradation. However, this reduction in power consequently lowers the DL sum rate.
Finally, Fig. \ref{fig4}(c) illustrates the impact of SI on the UL sum rate. It is observed that the performance gap between FD-RSMA-MA both side and FD-RSMA-MA BS side is relatively small compared to their gap with the other two schemes. This highlights the critical role of MAs at the BS in mitigating SI, as optimal antenna placement enables better SI suppression compared to FPAs. Consequently, schemes equipped with MA on the BS side demonstrate superior resilience to SI, maintaining a more stable sum rate despite increasing SI levels. At -60 dB of residual SI, our proposed FD-RSMA-MA both side achieves around 19.14\% performance improvement over FD-RSMA-MA BS side, 116\% improvement over FD-RSMA-MA UE side and 246.63\% improvement over FD-RSMA-FPA in terms of total sum rate respectively. Meanwhile, our proposed FD-RSMA-MA BS side achieves around 81.90\% performance gains over FD-RSMA-MA UE side and 190.94\% improvement over FD-RSMA-FPA at -60 dB of residual SI in terms of total sum rate. Finally, proposed FD-RSMA-MA UE side achieves around 59.94\% performance gains over FD-RSMA-FPA at -60 dB of residual SI in terms of sum rate. Moreover, it is noteworthy that FD-RSMA-MA both side can achieve feasible solution until -30 dB of residual SI while meeting all the constraints. Meanwhile, FD-RSMA-MA BS side and FD-RSMA-MA UE side stops giving feasible solutions after -45 dB. Finally, FD RSMA-FPA can produce feasible results until only -60 dB of residual SI. This phenomenon strongly reveals the MA capability in handling residual SI.
\begin{figure*}
\centering
\hspace{-0.35cm}\subfigure[] {\centering\includegraphics[width=0.35\textwidth]{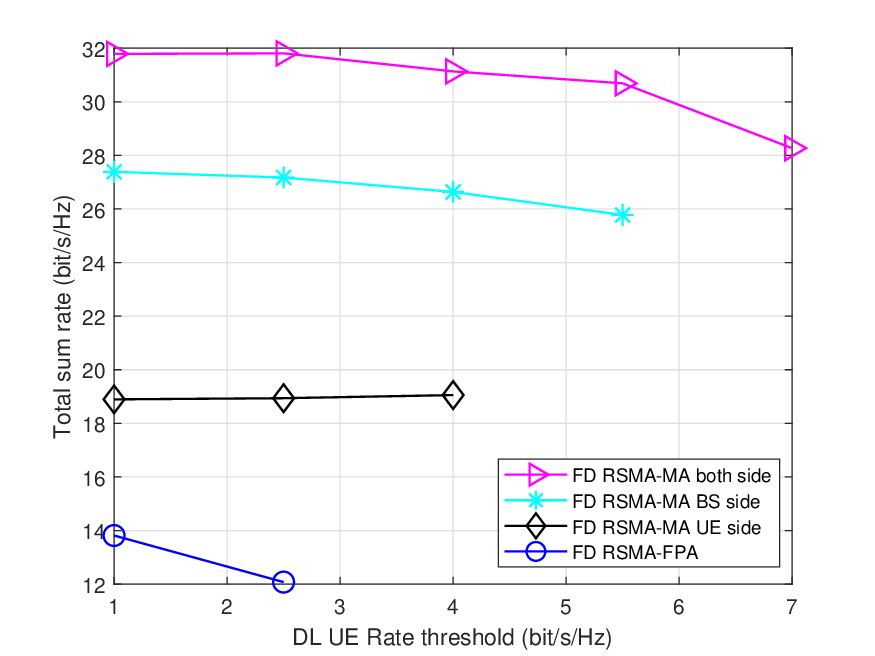}} 
\hspace{-0.5cm}\subfigure[] {\centering\includegraphics[width=0.35\textwidth]{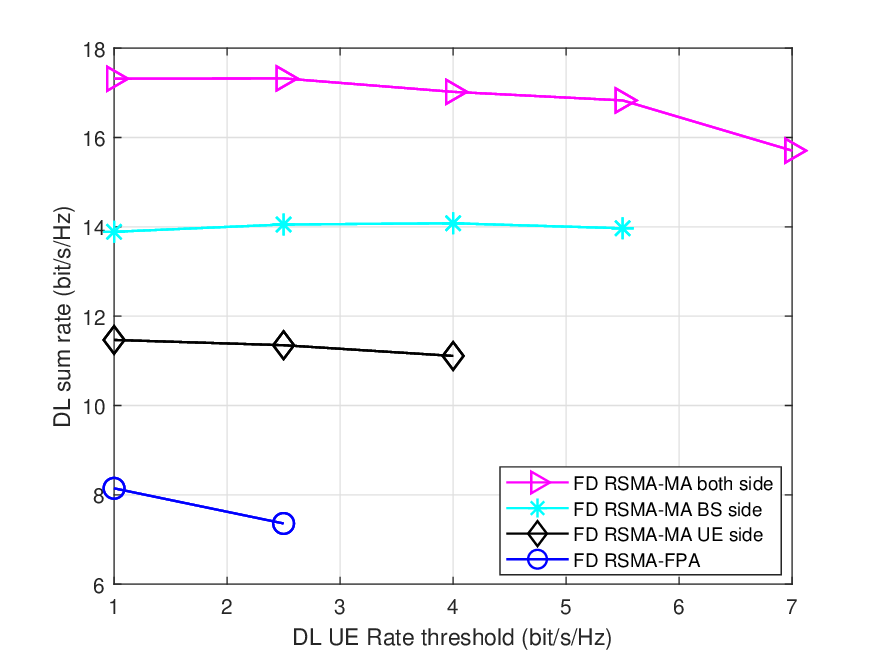}}  
\hspace{-0.5cm}\subfigure[] {\centering\includegraphics[width=0.35\textwidth]{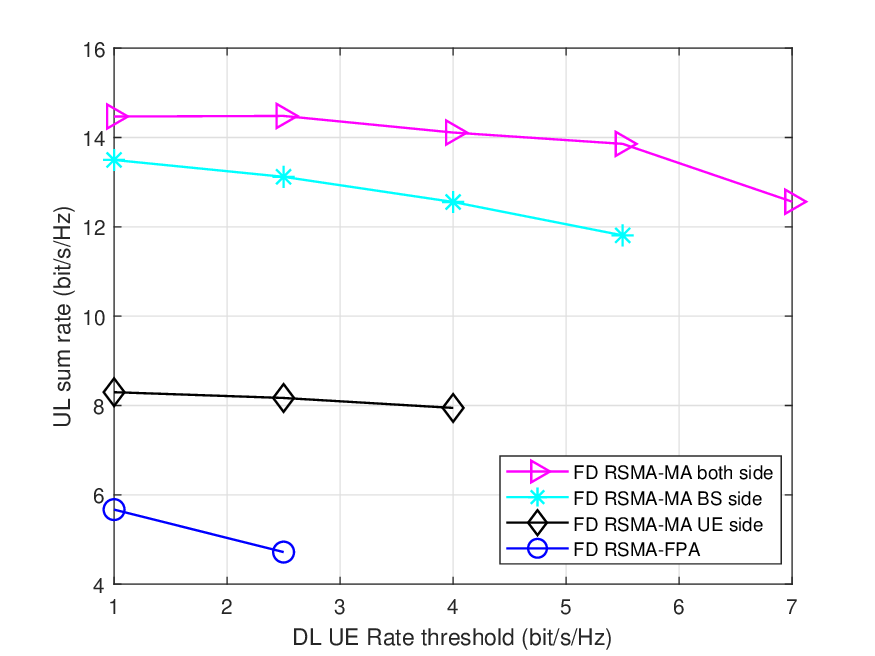}}
\caption{(a) Total achievable sum rate vs DL UE rate threshold (b) DL sum rate vs DL UE rate threshold (c) UL sum rate vs DL UE rate threshold}
\label{fig15}
\vspace{-.2 in}
\end{figure*}
%\vspace{-.1 in}
\subsection{Impact of DL UE Rate Threshold}
Fig. \ref{fig15} illustrates the impact of varying the DL UE rate threshold on the overall sum rate, as well as the individual UL and DL sum rates. As the DL rate threshold increases from 1 bps/Hz to 7 bps/Hz, the total sum rate across all schemes exhibits a declining trend. This reduction occurs because a larger portion of the BS's power budget is allocated to meet the increasing DL rate requirements, leaving fewer resources available for maximizing the sum rate.
Despite this general degradation in performance,  FD-RSMA-MA both side consistently achieves the highest sum rate among the evaluated schemes. However, it is important to note that FD-RSMA-MA both side reaches infeasibility beyond rate thresholds of 7 bps/Hz, while FD-RSMA-MA BS side ceases to provide feasible solutions beyond 5.5 bps/Hz. In contrast, the solutions of FD-RSMA-MA UE side becomes infeasible after exceeding 4 bps/Hz, and FD-RSMAA-FPA fails to meet feasibility constraints beyond 2.5 bps/Hz. These findings underscore the superior resilience of MAs in accommodating higher data rate demands compared to FPAs. Moving on to Fig.\ref{fig15}(b), we can see that it also supports our previous claim and shows similar trend as Fig.\ref{fig15}(a). Interestingly, Fig.\ref{fig15}(c) reveals that increasing the DL UE rate threshold also affects the UL sum rate. This phenomenon arises due to the adverse impact of residual SI on UL reception at the BS. Specifically, as the BS transmits at higher power to satisfy the increasing DL rate requirements, the residual SI at the BS intensifies, degrading its ability to efficiently decode UL signals. To mitigate this effect and maintain balanced overall system performance, the optimizer strategically reduces the UL sum rate, thereby limiting the SI impact on UL reception. As a result, the UL sum rate declines as the DL UE rate threshold increases.
%\vspace{-.2 in}
\section{Conclusion}
With the aim of enhancing the SE in terms of sum rate of FD-BS RSMA system by handling the negative impact of residual SI and intra-cell interference, we investigate the joint optimization of transmitting and receiving beamforming vectors at BS, UL UE transmission power, the common split portion of RSMA, and MA positions, which was formulated as a non-convex optimization problem. By invoking the GML technique, the non-convex optimization problem is solved in a distributed manner for the sake of the tractability of the problem. Finally, simulation results under various system parameter settings validate that deploying MA at an FD system can obtain more favorable outcomes than FPA by effectively combating residual SI and intra-cell interference. Particularly, our numerical results reveal that deploying MAs at BS is more effective than having MA only at the UE side, and the best outcome is achieved when MA is deployed at both the transmitter and receiver sides. 
\bibliographystyle{IEEEtran}
\bibliography{ref}

\end{document}